\documentclass[twocolumn,trackchanges]{aastex7}
\makeatletter
\providecommand{\@journalinfo}{}
\makeatother
\usepackage{amsmath}

\received{July 16, 2025}
\revised{January 8, 2026}
\accepted{January 12, 2026}

\begin{document}

submitjournal{ PASP}

\title{A High-frequency Geodetic VLBI Experiment for Optical Clock Comparison}

\correspondingauthor{Monia Negusini, monia.negusini@inaf.it} 

\author[0000-0002-0064-5533]{Monia Negusini}
\affiliation{Institute for Radio Astronomy, National Institute for Astrophysics, Via Piero Gobetti 101, Bologna, 40129, Italy}
\email{monia.negusini@inaf.it}

\author[0000-0002-9231-3789]{Myoung-Sun Heo}
\affiliation{Korea Research Institute of Standards and Science, 267 Gajeong-ro, Yuseong-gu, Daejeon, 34113, Republic of Korea}
\email{hms1005@kriss.re.kr}

\author[0000-0002-7289-6403]{Cecilia Clivati}
\affiliation{Quantum Metrology and Nanotechnology division, Istituto Nazionale di Ricerca Metrologica, Strada delle cacce 91, Torino, 10135, Italy}
\email{c.clivati@inrim.it}

\author[0000-0003-2953-6442]{Shuangjing Xu}
\affiliation{Korea Astronomy and Space Science Institute, 776 Daedeok-daero, Yuseong-gu, Daejeon, 34055, Republic of Korea}
\email{sjxu@kasi.re.kr}

\author[0000-0003-4631-1528]{Roberto Ricci}
\affiliation{Institute for Radio Astronomy, National Institute for Astrophysics, Via Piero Gobetti 101, Bologna, 40129, Italy}
\affiliation{Dipartimento di Fisica, Universit\'a di Roma Tor Vergata, Via della ricerca scientifica 1, Roma, 00133, Italy}
\email{roberto.ricci@inaf.it}

\author[0000-0001-7003-8643]{Taehyun Jung}
\affiliation{Korea Astronomy and Space Science Institute, 776 Daedeok-daero, Yuseong-gu, Daejeon, 34055, Republic of Korea}
\email{thjung@kasi.re.kr}

\author{Buseung Cho}
\affiliation{Korea Institute of Science and Technology Information, 245 Daehak-ro, Yuseong-gu, Daejeon, 34141, Republic of Korea}
\email{bscho@kisti.re.kr}

\author[0000-0003-0294-0365]{Matteo Stagni}
\affiliation{Institute for Radio Astronomy, National Institute for Astrophysics, Via Piero Gobetti 101, Bologna, 40129, Italy}
\email{matteo.stagni@inaf.it}

\author[0000-0002-6480-4564]{Claudio Bortolotti}
\affiliation{Institute for Radio Astronomy, National Institute for Astrophysics, Via Piero Gobetti 101, Bologna, 40129, Italy}
\email{claudio.bortolotti@inaf.it}

\author[0000-0002-1482-708X]{Giuseppe Maccaferri}
\affiliation{Institute for Radio Astronomy, National Institute for Astrophysics, Via Piero Gobetti 101, Bologna, 40129, Italy}
\email{giuseppe.maccaferri@inaf.it}

\author[0000-0002-8935-8142]{Federico Perini}
\affiliation{Institute for Radio Astronomy, National Institute for Astrophysics, Via Piero Gobetti 101, Bologna, 40129, Italy}
\email{federico.perini@inaf.it}

\author[0000-0003-4142-3897]{Mauro Roma}
\affiliation{Institute for Radio Astronomy, National Institute for Astrophysics, Via Piero Gobetti 101, Bologna, 40129, Italy}
\email{mauro.roma@inaf.it}

\author[0000-0003-1157-4109]{Do-Young Byun}
\affiliation{Korea Astronomy and Space Science Institute, 776 Daedeok-daero, Yuseong-gu, Daejeon, 34055, Republic of Korea}
\affiliation{University of Science and Technology, Gajeong-ro 217, Yuseong-gu, Daejeon, 34113, Republic of Korea}
\email{bdy@kasi.re.kr}

\author[0009-0003-6409-3218]{Do-Heung Je}
\affiliation{Korea Astronomy and Space Science Institute, 776 Daedeok-daero, Yuseong-gu, Daejeon, 34055, Republic of Korea}
\email{dhje@kasi.re.kr}

\author[0000-0003-2353-362X]{Marco Pizzocaro}
\affiliation{Quantum Metrology and Nanotechnology division, Istituto Nazionale di Ricerca Metrologica, Strada delle cacce 91, Torino, 10135, Italy}
\email{m.pizzocaro@inrim.it}

\author[0000-0002-0345-859X]{Davide Calonico}
\affiliation{Quantum Metrology and Nanotechnology division, Istituto Nazionale di Ricerca Metrologica, Strada delle cacce 91, Torino, 10135, Italy}
\email{d.calonico@inrim.it}

\author[0000-0002-2332-2534]{Elena Cantoni}
\affiliation{Quantum Metrology and Nanotechnology division, Istituto Nazionale di Ricerca Metrologica, Strada delle cacce 91, Torino, 10135, Italy}
\email{e.cantoni@inrim.it}

\author[0000-0002-4486-7138]{Giancarlo Cerretto}
\affiliation{Quantum Metrology and Nanotechnology division, Istituto Nazionale di Ricerca Metrologica, Strada delle cacce 91, Torino, 10135, Italy}
\email{g.cerretto@inrim.it}

\author[0000-0003-4921-7959]{Stefano Condio}
\affiliation{Quantum Metrology and Nanotechnology division, Istituto Nazionale di Ricerca Metrologica, Strada delle cacce 91, Torino, 10135, Italy}
\email{s.condio@inrim.it}

\author[0000-0002-7474-0349]{Giovanni A. Costanzo}
\affiliation{Quantum Metrology and Nanotechnology division, Istituto Nazionale di Ricerca Metrologica, Strada delle cacce 91, Torino, 10135, Italy}
\affiliation{Electronics and Telecommunications Department., Politecnico di Torino, Corso Duca degli Abruzzi 124, Torino, 10134, Italy}
\email{g.costanzo@inrim.it}

\author[0000-0002-9565-1214]{Simone Donadello}
\affiliation{Quantum Metrology and Nanotechnology division, Istituto Nazionale di Ricerca Metrologica, Strada delle cacce 91, Torino, 10135, Italy}
\email{s.donadello@inrim.it}

\author[0000-0002-4826-6495]{Irene Goti}
\affiliation{Quantum Metrology and Nanotechnology division, Istituto Nazionale di Ricerca Metrologica, Strada delle cacce 91, Torino, 10135, Italy}
\email{i.goti@inrim.it}

\author[0000-0001-8783-5665]{Michele Gozzelino}
\affiliation{Quantum Metrology and Nanotechnology division, Istituto Nazionale di Ricerca Metrologica, Strada delle cacce 91, Torino, 10135, Italy}
\email{m.gozzelino@inrim.it}

\author[0000-0002-0206-9082]{Filippo Levi}
\affiliation{Quantum Metrology and Nanotechnology division, Istituto Nazionale di Ricerca Metrologica, Strada delle cacce 91, Torino, 10135, Italy}
\email{f.levi@inrim.it}

\author[0000-0001-8922-4729]{Alberto Mura}
\affiliation{Quantum Metrology and Nanotechnology division, Istituto Nazionale di Ricerca Metrologica, Strada delle cacce 91, Torino, 10135, Italy}
\email{a.mura@inrim.it}

\author[0000-0001-8046-5710]{Matias Risaro}
\affiliation{Quantum Metrology and Nanotechnology division, Istituto Nazionale di Ricerca Metrologica, Strada delle cacce 91, Torino, 10135, Italy}
\email{m.risaro@inrim.it}

\author[0000-0002-1623-5520]{Huidong Kim}
\affiliation{Korea Research Institute of Standards and Science, 267 Gajeong-ro, Yuseong-gu, Daejeon, 34113, Republic of Korea}
\email{khd250@kriss.re.kr}

\author[0000-0002-2142-8343]{Won-Kyu Lee}
\affiliation{Korea Research Institute of Standards and Science, 267 Gajeong-ro, Yuseong-gu, Daejeon, 34113, Republic of Korea}
\email{oneqlee@kriss.re.kr}

\author[0000-0002-3139-6728]{Chang Yong Park}
\affiliation{Korea Research Institute of Standards and Science, 267 Gajeong-ro, Yuseong-gu, Daejeon, 34113, Republic of Korea}
\email{cypark@kriss.re.kr}

\author[0000-0001-7796-7664]{Dai-Hyuk Yu}
\affiliation{Korea Research Institute of Standards and Science, 267 Gajeong-ro, Yuseong-gu, Daejeon, 34113, Republic of Korea}
\email{dhyu@kriss.re.kr}

\author[0000-0003-2753-5227]{Young Kyu Lee}
\affiliation{Korea Research Institute of Standards and Science, 267 Gajeong-ro, Yuseong-gu, Daejeon, 34113, Republic of Korea}
\email{ykleeks@kriss.re.kr}

\author[0000-0001-7304-3624]{Joon Hyo Rhee}
\affiliation{Korea Research Institute of Standards and Science, 267 Gajeong-ro, Yuseong-gu, Daejeon, 34113, Republic of Korea}
\email{jh.rhee@kriss.re.kr}

\author[0009-0002-7557-9881]{Chanjin Park}
\affiliation{Korea Institute of Science and Technology Information, 245 Daehak-ro, Yuseong-gu, Daejeon, 34141, Republic of Korea}
\email{pcj0722@kisti.re.kr}

\author{Minseong Lee}
\affiliation{Korea Institute of Science and Technology Information, 245 Daehak-ro, Yuseong-gu, Daejeon, 34141, Republic of Korea}
\email{min0764@kisti.re.kr}

\author[0009-0004-4853-1019]{Hyo Ryoung Kim}
\affiliation{Korea Astronomy and Space Science Institute, 776 Daedeok-daero, Yuseong-gu, Daejeon, 34055, Republic of Korea}
\email{hrkim@kasi.re.kr}

\author[0000-0001-6280-8222]{Sung-Moon Yoo}
\affiliation{Korea Astronomy and Space Science Institute, 776 Daedeok-daero, Yuseong-gu, Daejeon, 34055, Republic of Korea}
\email{yoo@kasi.re.kr}

\author[0000-0002-3482-1232]{Jungho Cho}
\affiliation{Korea Astronomy and Space Science Institute, 776 Daedeok-daero, Yuseong-gu, Daejeon, 34055, Republic of Korea}
\email{jojh@kasi.re.kr}

\author[0000-0002-1229-0426]{Jongsoo Kim}
\affiliation{Korea Astronomy and Space Science Institute, 776 Daedeok-daero, Yuseong-gu, Daejeon, 34055, Republic of Korea}
\affiliation{University of Science and Technology, Gajeong-ro 217, Yuseong-gu, Daejeon, 34113, Republic of Korea}
\email{jskim@kasi.re.kr}

\author{Sang-Oh Yi}
\affiliation{National Geographic Information Institute, 92 Worldcup-ro, Yeongtong-gu, Suwon-si, 443-772, Republic of Korea}
\email{sangoh.yi@gmail.com}

\author[0000-0001-9917-5246]{Ha Su Yoon}
\affiliation{National Geographic Information Institute, 92 Worldcup-ro, Yeongtong-gu, Suwon-si, 443-772, Republic of Korea}
\email{hasuyoon@korea.kr}

\author[0000-0002-5902-5005]{Pablo de Vicente}
\affiliation{ Yebes Observatory, Instituto Geográfico Nacional, Cerro de la Palera, Yebes, Guadalajara, 19143, Spain}
\email{p.devicente@oan.es}

\author{Javier González}
\affiliation{ Yebes Observatory, Instituto Geográfico Nacional, Cerro de la Palera, Yebes, Guadalajara, 19143, Spain}
\email{j.gonzales@oan.es}

\author{Cristina García Miró}
\affiliation{ Yebes Observatory, Instituto Geográfico Nacional, Cerro de la Palera, Yebes, Guadalajara, 19143, Spain}
\email{c.garciamiro@oan.es}

\begin{abstract}

An intercontinental metrological clock comparison between Italy and the Republic of Korea was performed by means of geodetic K-band VLBI observations. The comparison involved the hydrogen masers (H-masers) used at Medicina and Sejong radio telescopes. The same clocks were simultaneously compared by a satellite link and by high-precision optical clocks maintained at the National Metrology Institutes, KRISS in Korea and INRIM in Italy, and delivered to VLBI antennas via optical fiber. 
The H-masers frequency difference was estimated by extrapolating the clock rate from VLBI data using two geodetic VLBI software. This was subsequently compared with clock differences derived by satellite link and by local optical clocks. Results obtained with different approaches were in agreement at the level of $10^{-15}$ s/s. This pilot study demonstrates that standard high-frequency (K-band) geodetic VLBI campaigns could be a viable approach to conduct intercontinental clock comparisons, now only possible via satellite links. This uncertainty can be reduced thanks to the planned installation of new-generation, broadband, high-frequency receivers on the involved telescopes. K/Q/W-band geodetic observations will allow an improvement of the accuracy of the resulting group delays through broad bandwidth synthesis from 20 to 100 GHz. Furthermore, the Frequency Phase Transfer (FPT) method will also be explored together with the use of PCAL systems installed at the radio telescopes to improve phase stability and thus allow a better estimation of the station clock parameters.

\end{abstract}



\section{Introduction}\label{sec:intro}

Very Long Baseline Interferometry (VLBI) is one of the scientific fields that most heavily relies on accurate time and frequency reference signals, demonstrating significant synergy with fundamental metrology. It is based on the  simultaneous observations of radio sources  with an array of telescopes, each referenced to a local  frequency standard. By correlating the radio signals received by the various telescopes it is possible to reconstruct pair-wise propagation delays, that depend on the baseline length and orientation, atmospheric effects and ultimately the radio source position and structure \citep{Schuh2012}. Discrepancies in the local clock frequencies and instrumental delays may also play a role. When all effects are taken into account, it is possible for VLBI  to reconstruct with good fidelity  rich information, e.g., on  the position of the radio telescopes, the Earth Orientation Parameters (EOPs), the source positions and structure.
Specifically, VLBI and frequency metrology can take mutual advantage from each other: on one side, VLBI resolution could be greatly improved with instrumentation and techniques that are traditionally maintained at Metrological Institutes. These include atomic clocks with state-of-the-art accuracy, today at the $10^{-18}$ level \citep{Beloy2021,Brewer2019,McGrew2018,Ushijima2015}, useful to investigate phenomena over timescales of months or years \citep{venka2020}; high spectral purity  oscillators \citep{Nakamura2020,xie2017}, crucial in mm- and sub-mm-wavelength observations \citep{raymond2024}; sub-ns  synchronization of remote sites \citep{Dierikx2016,Serrano2009} and distribution of atomic clock signals to multiple telescopes via optical fiber, that enable to realize distributed common-clock arrays \citep{krehlik17, Clivati2020, Boven2026}. On the other side, atomic clocks connected to the telescope can be compared worldwide via VLBI. This is  important for fundamental metrology, as discussion is ongoing on a future redefinition of the second in the International System of units \citep{Gill2016,Lodewyck2019,Riehle2018,dimarq2024}. Comparing distant atomic clocks is among the most urgent tasks to gain confidence on the best route to redefinition. Moreover, it allows fundamental physics tests at unprecedented resolution \citep{Sanner2019,  Lange2021}, as well as advanced relativistic geodesy measurements \citep{Grotti2018,McGrew2018,Takamoto2020}. 
However, clock comparisons represent a challenging task: 
while regional distances can be covered with fiber optic links \citep{Musha2008,Akatsuka2020,Lisdat2016,Droste2013,Clivati2020, Clivati2022topa,Husmann2021,Chiodo2015}, intercontinental clock comparisons are mostly conducted using the Global Navigation Satellite Systems (GNSS), though their performance may not meet the accuracy required for the redefinition \citep{Bauch2005a, Pizzocaro2021,Riedel2020,Leute2016,Fujieda2014,Hachisu2014}.

Since the end of the 1970s, the geodetic VLBI has been using the S/X band receivers with the aim of monitoring the parameters of the Earth's orientation, studying the movements of the Earth's crust and other geophysical phenomena, realizing the international terrestrial and celestial reference frames (ITRF and ICRF).
Its application for time and frequency transfer has been investigated since the beginning (e.g. \cite{Counselman1977, Clark1979, Hurd1979}) and an uncertainty of 1.5 × $10^{-15}$ for a time period of 1 day has been reported in an earlier study \citep{Rieck2012}. For local H-masers, which have a relative frequency instability well below $10^{-12}$, the VLBI-determined clock rate is equivalent to the rate of one clock relative to that of the other clock \citep{Sekido2021}. 
The uncertainties that can be achieved with legacy S/X band observations are not sufficient for optical clocks comparison, so new possibilities are being explored, such as VLBI observations performed with broadband receivers. 

In 2018/19, an intercontinental comparison of optical clocks was carried out using a broadband VLBI link between Italy and Japan with an uncertainty of $2.8  \times 10^{-16}$, lower than that achievable by a satellite link and thus particularly promising in a metrological perspective \citep{Pizzocaro2021}. However, in that campaign data were collected and analyzed under very special constraints, such as small transportable antennas and broadband NINJA feeds (3-14 GHz) and a direct digitization without baseband conversion \citep{Sekido2021}.
A new generation of a high-frequency broadband receiver ($>$ 20 GHz) is beginning to become available globally among VLBI communities, which would be highly effective in calibrating tropospheric phase fluctuations in the millimeter waveband, significantly enhancing the precision of group delay measurements through broad bandwidth synthesis from 20 to 100 GHz.   Moreover, they have the potential to reduce frequency-dependent systematic errors, such as source structure ones. 
The visible source structure on larger scales induces non-closure delays, increasing post-fit delay residuals in geodetic solutions, while the invisible structure within the beam size leads to variations in the source position \citep{2025arXiv250118276X}. Given the influence of in-beam structure on source positions, the transition to higher radio frequencies may be important for astrometry and geodesy.  
To fully exploit its potential, a highly stable local frequency reference is required as stated in \citep{rioja2020} and \citep{ledbetter25}. Rather than installing a new local reference at each site, accurate clocks delivered via optical fiber links offer a promising alternative.
These advantages can be beneficial for accurate clock comparison using VLBI.

A project involving the Korea Astronomy and Space Science Institute (KASI) and the Italian Institute of Astrophysics (INAF) allows the installation of Korean Compact Triple-band Receivers (CTRs) that operate simultaneously in the K (18-26 GHz), Q (35-50 GHz), and W (85-115 GHz) bands \citep{Han2017} on all three INAF radio telescopes (Medicina, Noto, and SRT). Waiting for the installation and commissioning phases to be completed, we explore the use of an intercontinental network of operational high-frequency VLBI antennas to conduct metrological clock comparisons. Importantly, our strategy was to exploit procedures and instrumentation that are routinely used at most radio telescopes today. This would leverage the use of the existing VLBI network for various scientific tasks, including metrology. 
Our test campaigns were thus conducted as geodetic ones, with K-band observations and standard scheduling and analysis, with the aim to identify ultimate limits, critical aspects and achievable metrological performances. 

This paper will describe the international network, experimental setup and analysis procedures, and discuss current results and future perspectives. Specifically in Section~\ref{sec:set} the basics of geodetic VLBI are presented and the VLBI-clock comparison experiment is explained, in Section~\ref{sec:obs} the VLBI observations and geodetic data analysis are presented. In Section~\ref{sec:res} the comparison via VLBI, optical clocks and GNSS are shown; in Section~\ref{sec:disc} the results are discussed and some conclusion and outlook are drawn in Section~\ref{sec:conc}.

\section{VLBI-clock comparison experiment} \label{sec:set}

\begin{figure*}[htb!]
\centering
\includegraphics[width=\textwidth]{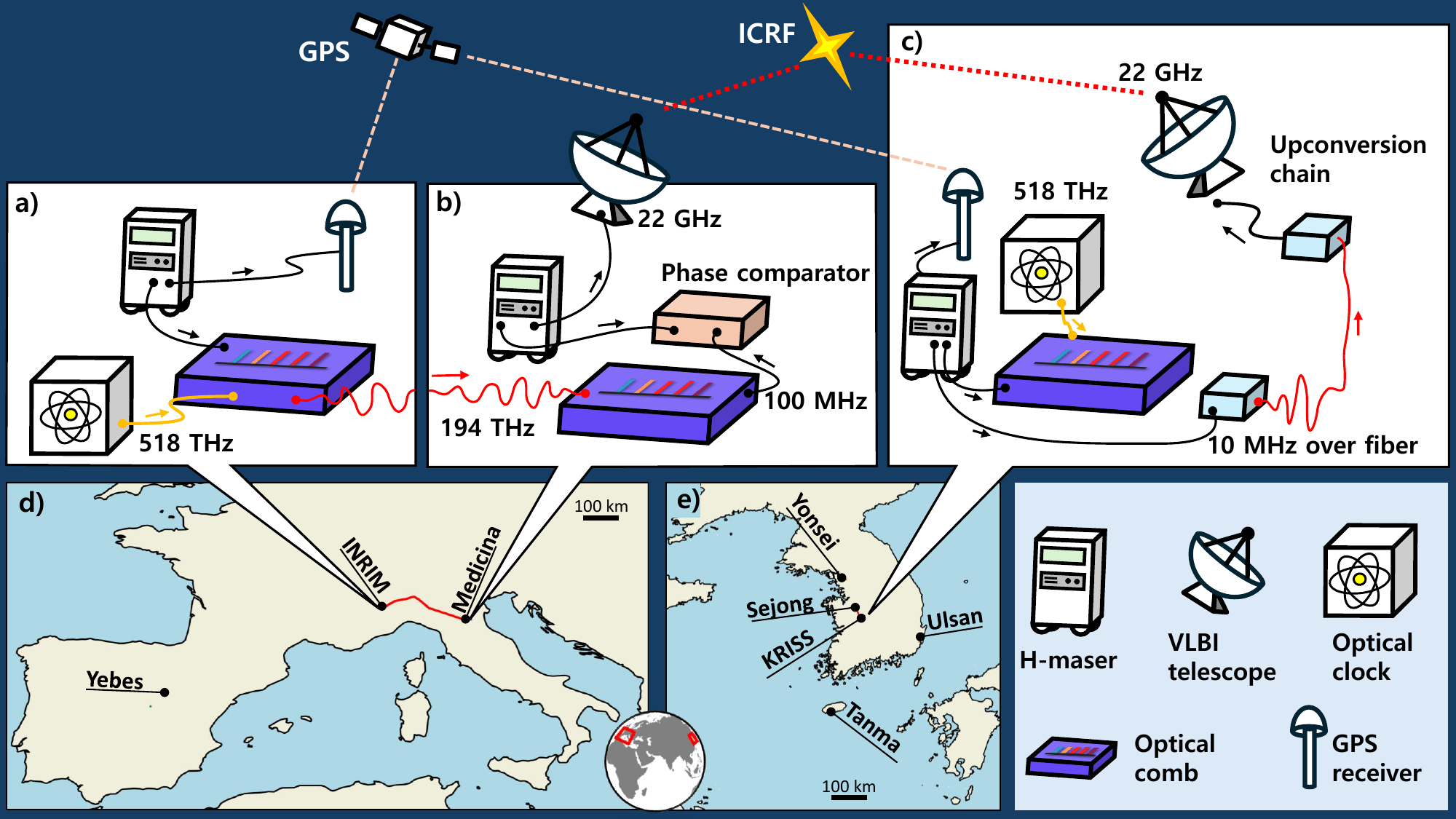}
\caption{Schematic view of the full experiment setup. a) At INRIM, in Italy, radiation from an infrared laser is referenced to IT-Yb1 clock via an optical comb and sent to Medicina with a phase-stabilised fiber. IT-Yb1 was also compared to KRISS-Yb1 via a GPS link, using INRIM H-maser as flywheel oscillator.  b) In Medicina, the incoming radiation is coherently converted to a radiofrequency and used to calibrate the local H-maser, which is used for the VLBI observation. c) In Korea, a 10 MHz signal referenced to a H-maser is sent  to Sejong VLBI antenna with a phase-stabilised fiber; here, after demodulation, it  seeds the antenna synthesis chain and is used for VLBI observation. KRISS H-maser is also calibrated by KRISS-Yb1 and compared to INRIM via  GPS. d) and e) Map of Europe and Korea with involved telescopes and Metrological Institutions.}\label{fig:fig1}
\end{figure*}

\subsection{VLBI basics} 

Geodetic VLBI is based on the simultaneous observation of extragalactic radio sources (usually quasars) with distant antennas. The main observable is the difference in arrival times (time delay) of the radio signal from a radio source to two or more globally distributed radio telescopes. This time delay $\tau_{\text{obs}}$, measured to a precision of a few picoseconds using highly stable atomic clocks (hydrogen masers) at each station, is the fundamental measurement used to derive essential geodetic and astrometric parameters, including precise positions and velocities of the telescopes, positions of quasars, and Earth orientation parameters, such as Universal Time (UT1), which describes the Earth's rotation angle.
Indeed, the geometric delay between two stations, $\tau_{\text{geom}}= \textbf{B}\cdot\textbf{S}/\textit{c}$, depends on the baseline vector between the antennas (\textbf{B}) and the unit vector of the radio source (\textbf{S}), where \textit{c} is the speed of light.
The observed value $\tau_{\text{obs}}$ differs from the geometric delay due to a number of different effects, such as the time offset between the reference clocks $\Delta\tau_{clock}$, the station-to-station difference in the excess delays introduced by the neutral (troposphere, $\Delta\tau_{\text{tro}}$) and the ionized (ionosphere, $\Delta\tau_{\text{ion}}$) part of the atmosphere, instrumental delays such as antennas, cables and receivers $\Delta\tau_{\text{instr}}$ and finally the structure of the radio source $\Delta\tau_{\text{source}}$:
\begin{equation}
    \tau_{\text{obs}} = \tau_{\text{geom}} + \Delta\tau_{\text{tro}} + \Delta\tau_{\text{ion}} + \Delta\tau_{\text{instr}} + \Delta\tau_{\text{source}} + \Delta\tau_{\text{clock}}
     \label{eq:eq1}
\end{equation}
The different components of the delay can be modelled, estimated from the observations or obtained as a result of external measurements.
Typically, geodetic VLBI experiments last $24$ hours, involving a network of stations and observing a number of spatially distributed radio sources with very well known positions in order to have a good coverage of the sky and thus be able to estimate with very good accuracy the positions of the radio telescopes, the atmospheric delays and the clock behavior.

\subsection{Experiment setup}

In this work, we compare the frequencies of H-masers used in VLBI observations at Medicina and Sejong radio telescopes via a VLBI geodetic link by measuring $\Delta \tau_{\text{clock}}$ as described in the previous section and, simultaneously, by calibration to high-precision Yb optical clocks, KRISS-Yb1 and IT-Yb1, maintained at the National Metrology Institutes (NMIs) Korea Research Institute of Standards and Science (KRISS) in Korea and Istituto Nazionale di Ricerca Metrologica (INRIM) in Italy, respectively (Fig.~\ref{fig:fig1}).
During the VLBI sessions, a GNSS link based on GPS Precise Point Positioning (PPP) was operational between INRIM and KRISS for consistency check and comparison.

Throughout this work, the quantity directly compared by VLBI is the fractional frequency difference between the two H‑masers; the optical clocks and the GPS link are used to calibrate these masers and to provide an external reference for evaluating the VLBI link itself.

\subsection{VLBI setup}          
A network of six radio telescopes between Europe and the Republic of Korea was involved in the K-band 24-h geodetic session: Medicina 32-m antenna (Italy, operated by INAF) and Yebes 40-m antenna (Spain, operated by the Yebes Astronomical Centre) in Europe (Fig. \ref{fig:fig1}d), and Sejong 22-m antenna (operated by the National Geographic Information Institute - NGII) and the three Yonsei, Ulsan and Tamna 21-m Korean VLBI Network antennas (KVN, operated by KASI), in Korea (Fig. \ref{fig:fig1}e). The Yebes antenna was involved in the experiment to balance the geometry of the network between Europe and Korea and to provide additional observations to improve the overall statistics of the observing session and the estimation of interesting parameters, by increasing the number of baselines. 
The Sejong station receives a clock signal (10 MHz) from a hydrogen maser at KRISS via the optical fiber of the Korea Research Environment Open Network (KREONET), a national research and science network operated by the Korean Institute of Science \& Technology Information (KISTI) (Fig. \ref{fig:fig1}c). The delivered signal is demodulated and multiplied to 1.4 GHz, to feed into the Round Trip System (RTS) \citep{Oh2010}, which sends the 1.4 GHz signal to the antenna receiver room for VLBI observation with the fiber-induced noise suppressed. At Medicina station, in Italy, the local H-maser is used for VLBI observations, but its frequency is constantly measured against a clock signal delivered from INRIM over the Italian Quantum Backbone (IQB), a fiber infrastructure for metrological time and frequency distribution in the whole country (Fig. \ref{fig:fig1}a and \ref{fig:fig1}b). More details on the fiber-based clock distribution to the two telescopes are given in Section~\ref{sec:ocfd}.
The experiment schedule was generated using NASA's software {\it SKED} \citep{Gipson2010},
which is widely used to schedule geodetic and astrometric  VLBI observations.
We compiled a catalog with over 230 sources, which are included in the ICRF3 K-band catalog \citep{charlot20} and generally strong ($>0.4$ Jy at K or X band), flat-spectrum sources in the Radio Fundamental Catalog \citep{Petrov2024}.
The best 69 targets, ranked primarily by sky coverage, were automatically selected by {\it SKED}  for the six-station network during the observation.
The sequence of source scans is well-separated in hour angle and elevation over time to improve astrometric accuracy and help separate the tropospheric effect from other parameters.
A total of 451 scans were scheduled, each with an integration time of 120 seconds per source. We recorded right circular polarization (RCP) at a frequency range between 21184 and 21696 MHz with a total data rate of 2048 Mbps. The total bandwidth of 512 MHz was divided into 16 intermediate frequency (IF) bands. 


\subsection{Optical clocks and fiber distribution}\label{sec:ocfd}
      
To calibrate H-masers, we used KRISS-Yb1  and IT-Yb1, two Yb optical lattice clocks both with systematic uncertainty of $2 \times 10^{-17}$,  \citep{Goti2023, Kim2021}. Both clocks contribute regularly  to the generation of the International Atomic Time (TAI). 
INRIM and KRISS are also equipped with GNSS receivers for time transfer. 
The main reference oscillator at KRISS is a H-maser (KRISS-HM) traceable to UTC(KRIS); its frequency is calibrated by KRISS-Yb1 via an optical frequency comb. 
The 10 MHz signal from the hydrogen maser is transferred to the Sejong VLBI antenna via a 50 km fiber link. We used a commercial fiber optic frequency distribution system (OSTT, PikTime) to actively compensate for fiber noise (Fig. \ref{fig:fig1}c). 

The additive noise resulting from the fiber link was measured in KRISS to be $10^{-13}$ at 1 s and $2\times10^{-16}$ at $10^{4}$ s, so the fiber link did not degrade the performance of the VLBI measurement. 

In Italy, the Medicina telescope is connected to INRIM via a 535 km optical fiber which is part of the IQB. Among other research facilities, this infrastructure connects two of the main Italian radio telescopes,  Medicina and Matera, and has already been used to carry out a common-clock VLBI experiment \citep{Clivati2020}. 
The frequency distribution chain is based on an ultrastable laser, stabilized at INRIM to a high-finesse Fabry-Perot cavity and calibrated by IT-Yb1 via an optical comb (Fig. \ref{fig:fig1}a). The optical signal is sent to Medicina and here used to reference a second optical comb, from which an ultra-low noise microwave at 10 GHz and 100 MHz is extracted and phase compared to the local H-maser (Medicina-HM) (Fig. \ref{fig:fig1}b). The fiber is stabilized using the Doppler noise stabilization technique, so that the frequency signal delivered at the telescope has a stability of a few parts in $10^{-14} $ at 1 s and $<10^{-16} $ at $10^4$ s \citep{Clivati2015}. 
                    
\section{VLBI observation and data analysis} \label{sec:obs}

\subsection{Observation overview} \label{sec:cam} 
The 24-hour VLBI experiment was carried out at 6 stations from UTC 2021-Dec-16 19:00 to 2021-Dec-17 19:00. The weather conditions were different between Europe and the Republic of Korea. The weather was good in Medicina and Yebes while snow and strong wind were reported at the Sejong station. 
The KVN observations were completed without any major problems.
Approximately, the last seven hours of Yebes data at the end of the experiment were not recorded, thus affecting the last part of the experiment. 

The optical clock and fiber-based distribution systems operated almost continuously from a few hours before to a few hours after the campaign. During the 24-hour-long campaign IT-Yb1 and KRISS-Yb1 had  an  uptime of 94\% and 97\% respectively. 
The GNSS link was  operational during the campaign without interruption.

\subsection{VLBI data correlation and fringe fitting}\label{sec:vlbicorr}

Data was transferred from each station to the Daejeon and Bologna Correlators and here processed to compute pairwise interference fringes (fringe-fitting) and extract relative delays between pairs of antennas ($\tau_{\text{obs, i,j}}$). In detail, we used the Distributed FX (DiFX) correlation software \citep{Deller2007}, by performing fast Fourier transform (FFT) followed by a complex multiplication of the raw visibilities data of each antenna taken in pairs. Prior to this, the estimation of parameters such as station positions and velocities, source coordinates, and Earth orientation parameters were calculated to achieve the best possible delay model using the Calc software \citep{Gordon2017}.

After the correlation was completed, the DiFX files were converted into Mark4 directories and then fringe-fitted using Haystack Observatory Postprocessing System {\it fourfit} \citep{Hoak2022}. In the fringe fitting procedure the peak intensity as a function of delay and delay rate is computed scan by scan and sub-band by sub-band and then applied to the fringe visibility data in order to flatten the phases in the frequency channels in each sub-band. Manual phase calibration and Lower Side Band (LSB) offset were inserted for all stations before running the fitting routine.

After the fringe fitting was applied, the data appeared in the Mark3 format, but the commonly used software programs for geodetic data analysis work with databases. For this reason the Mark3 fringe fitting output files were converted into vgosDb \citep{Bolotin2017} databases containing observed delays, {\it a-priori} models, instrumental delays corrections and weather parameters. 

\subsection{Geodetic data analysis} \label{sec:geoan}

Pairwise delays derived as described in Sec. \ref{sec:vlbicorr} were then fitted with a global model to extract quantities reported in Eq. \ref{eq:eq1}.
The data analysis software $\nu$Solve \citep{Bolotin2014} was used to analyze the vgosDb database, named 21DEC16XF. $\nu$Solve solves for the normal equation calculating the adjustments of a set of parameters by comparing model group delays with observed ones on antenna pairs for each source scan in an observing session. The software estimates the clock parameters through a quadratic model (offset, clock rate, and quadratic term, which corresponds to the relative frequency drift) plus a piece-wise linear function. 
All these clock parameters are estimated with respect to a reference clock, in this work the Medicina H-maser.
Tropospheric parameters, station positions and EOPs are also estimated in the single session solution.

For verification of any possible systematic errors in the analysis, an additional analysis was conducted using the Vienna VLBI \& Satellite Software for astrometry and geodesy (VieVS) \citep{Boehm2018}, developed by the Vienna Geoinformatics Group. 
Unlike $\nu$Solve, which performs ionospheric correction only in dual-band sessions (e.g. S/X-band), VieVS is able to perform ionospheric delay correction even in single-band sessions (e.g. K-band). It utilizes GNSS-generated Global Ionospheric Maps (GIMs) to compute ionospheric correction at the position of each station at the time of  each observing scan in the session. Such maps have a latitude/longitude resolution of 2.5/5.0 degrees and are provided daily by the International GNSS Service (IGS) through the Crustal Dynamics Data Information System (CDDIS) repository \citep{CDDIS}. 
The corrections obtained from this external model were applied to the observing session database.

\section{Results}\label{sec:res}

\subsection{Comparison via VLBI}

\begin{figure}[htb!]
\centering
\includegraphics[width=0.5\textwidth]{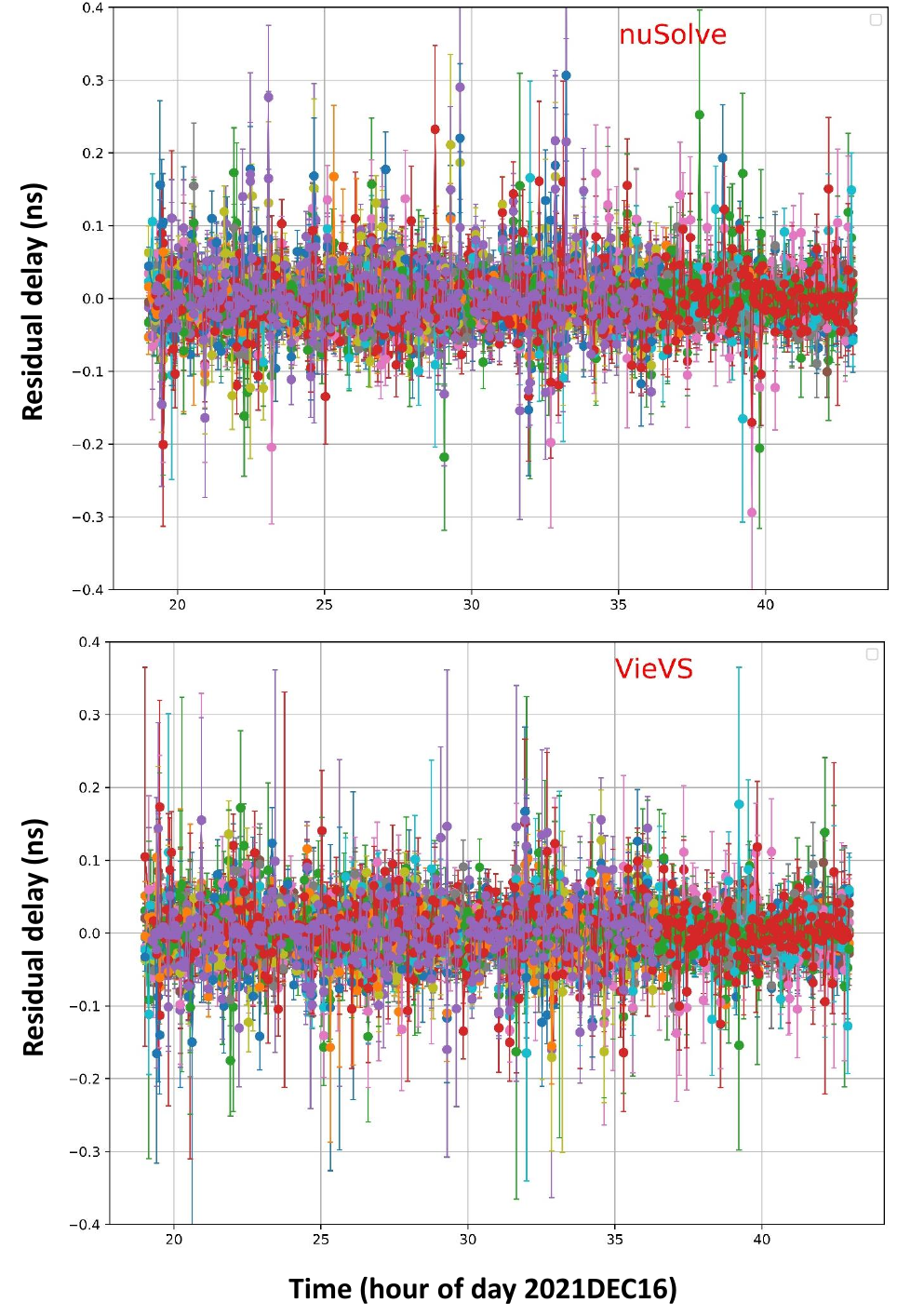}
\caption{Group delay residuals as a function of observing time during the Dec 2021 geodetic session obtained by $\nu$Solve (top) and by VieVS (bottom). Different colors represent the results of distinct baselines.}
\label{fig:res}
\end{figure}

$\nu$Solve was run on the 21DEC16XF database using standard settings for the parametrization, in order to estimate the clock model parameters (CL0: offset; CL1: clock rate and CL2: quadratic term) needed for clock comparison.
As stated in Section~\ref{sec:obs} no ionospheric corrections were applied as the dataset is single-band. 
After the least-square minimization using 5793 observation pairs, a weighted root-mean-square (WRMS) of the group delay residuals of 25.55 ps was obtained, which is the difference between the observed time delay (measured by the radio telescopes) and the theoretical delay (calculated based on known models) and quantifies the overall quality of the experiment. The group delay residuals as a function of observing time are shown in Fig.~\ref{fig:res} (top). The estimates of the second term CL1 and third term CL2 of the clock model are reported in Table~\ref{tab:clk}.

VieVS was run on the same database using standard settings for the parametrization. The main differences with $\nu$Solve concerned the use of the VMF3 mapping function \citep{Landskron2018} for the troposphere parameters estimation, and of the {\it a priori} GPT3 model \citep{Landskron2018} for estimating their azimuthal gradients.
Two VieVS solutions were obtained with and without the application of the ionospheric correction, respectively. There was no significant difference between the two solutions. 
The bottom plot in Fig.~\ref{fig:res} shows the group delay residuals with the ionospheric correction applied. Using 5784 observation pairs, we obtained a WRMS of 26.41 ps.
The estimates of the clock rate CL1 and quadratic term CL2 of the clock model are reported in Table~\ref{tab:clk}. Specifically, we assumed CL1 as evaluated with VieVS to be the fractional frequency difference $y$ between the HMs in Medicina and KRISS obtained by VLBI, $y$(KRISS-HM - Medicina-HM, VLBI)=$-212.5(1.1)\times 10^{-15}$. 

We adopted clock parameters obtained using VieVS as the best estimation because they were obtained using a better tropospheric and ionospheric modeling. However, the two methods were consistent within uncertainty.\\

\begin{deluxetable}{lcc}
\tablecaption{
Relative difference in frequency CL1 and its drift CL2 of the H-maser at Sejong from that at Medicina obtained from the geodetic analysis performed on $\nu$Solve and VieVS.}
\label{tab:clk}
\tablewidth{0pt}
\tablehead{
\colhead{Tool} & \colhead{CL1} & \colhead{CL2} \\
\colhead{} & \colhead{   ($1\times 10^{-15}$) s/s} & \colhead{ ($1\times 10^{-15}$) day$^{-1}$}
}
\startdata
$\nu$Solve  & -214.3 (1.5) &   2.1 (2.3)    \\
VieVS       & -212.5 (1.1) &  -0.7 (1.0)  \\
\enddata
\end{deluxetable}

\subsection{Comparison by optical clocks}

H-masers used for the VLBI measurements at Sejong and Medicina were calibrated by KRISS-Yb1 and IT-Yb1.  
By calibration we mean that the absolute frequencies of the two masers are derived from the comparison with optical clocks, as the two Yb clocks are secondary representations of the SI second \citep{Margolis2024}. In calculating the frequency difference between two H-masers we will assume that, being based on the same atomic species, the two Yb clocks have the same frequency within their uncertainty. 
In this way, we calculated the average relative frequency difference $y$ of the masers during the campaign, determined by optical clocks (OC), as $y$(\text{KRISS-HM - Medicina-HM, OC}) =  $-213.29(14) \times 10^{-15}$. Here the uncertainty includes the instability ($7\times 10^{-18}$ and  $1.1 \times 10^{-17}$) and the systematic uncertainty of the clocks ($2\times10^{-17}$ and  $2\times 10^{-17}$) respectively for IT-Yb1 and KRISS-Yb1. The uncertainty includes an extrapolation uncertainty of $4\times10^{-17}$ for Medicina-HM and $7\times 10^{-17}$ for KRISS-HM due to the dead-time in the optical clock measurements calculated from the known noise of the two masers \citep{Yu2007, Hachisu2015}. This approach is common for optical clocks operating intermittently \citep{Grebing2016, Leute2016, Hachisu2017, Pizzocaro2020, Pizzocaro2021, Nemitz2021}. The frequency conversion of the combs introduces an uncertainty of $5\times10^{-17}$ in Medicina and $11\times 10^{-17}$ in Sejong. The optical fiber link connecting IT-Yb1 in Torino to the H-maser in Medicina has been characterized to better than $1\times 10^{-18}$ and contributes negligibly to the uncertainty of the measurement.
We note that in December 2021 the optical clocks IT-Yb1 and KRISS-Yb1 provided an extended set of data to contribute to the realization of TAI. IT-Yb1 contributed 5 days of data (MJD 59564-59569) that appeared in Circular T n. 410. KRISS-Yb1 contributed 35 days of data (MJD 59544 59579) that appeared in Circular T n. 408. A direct comparison of the two measurements shows the agreement of the two optical clocks with the relative frequency difference of $y$(\text{IT-Yb1 - KRISS-Yb1}) =  $3(12) \times 10^{-16}$, where the uncertainty ($1.2 \times 10^{-15}$) was limited by satellite link over the 5-day measurement time of IT-Yb1 as appeared in Circular T, and where we included an extrapolation uncertainty using the TAI stability to consider the different measurement time of the two clocks \citep{Pizzocaro2021}. Still, consistency in the two clocks at the $10^{-17}$ level is expected from the two uncertainty budgets.

\subsection{Comparison via GNSS}
A GPS-based independent and calibrated time and frequency transfer link was established to validate the VLBI intercontinental link, using two geodetic receivers for timing applications at INRIM and KRISS. These receivers track the code and phase signals emitted by the GPS satellites, and allow comparing them with the internal receiver clock, in turns synchronized and frequency locked to the external signal that is object of comparison. 
At INRIM, two Septentrio PolaRx4-TR receivers were used, connected to UTC(IT) and the H-maser at INRIM; at KRISS, a GTR55 receiver (MESIT asd) was used to connect to KRISS H-maser which is the master clock for KRISS-Yb1.
Among various possible processing algorithms, we chose the PPP \citep{kouba} as it includes the compensation of a significant set of atmosphere, geodynamics, and satellite effects \citep{kouba}.  
Intending PPP as a way of processing, different implementations are available worldwide. We chose NRCan PPP, produced by Natural Resources Canada and optimized for timing applications also with contribution of INRIM, now in use for more than 10 years by the BIPM for the computation of the TAI and UTC international time scales.
Precise IGS estimations for GPS satellites' orbits and clocks were retrieved from the IGS repositories.

The PPP solution was obtained by processing data for the period including the VLBI campaign from MJD 59562 to MJD 59571 and  by combining them with data from the optical-VLBI link. 

We calculate the average difference of the maser frequency during the campaign using GPS as  $y$(\text{KRISS-HM - Medicina-HM, GPS}) = $ -215.7(2.0) \times 10^{-15}$. 

The uncertainty is limited by the averaging time of the PPP solution over 1 day. We note that other campaigns using GPS data at INRIM \citep{Clivati2022topa} revealed a possible problem with the GPS solution at the level of $3 \times 10^{-16}$ in the considered period of time. This effect is negligible for the measurement presented here.

\begin{deluxetable*}{lcc}
\tablecaption{Average frequency difference in relative units of the H-masers at Sejong and Medicina during the campaign, measured using VLBI, the GPS link and the calibration by optical clocks.}
\label{tab:hmdiff}
\tablewidth{0pt}
\tablehead{
\colhead{} & \colhead{Maser frequency difference} & \colhead{Uncertainty} \\
\colhead{} & \colhead{$y$  ($1 \times 10^{-15}$)} & \colhead{$u_\text{tot}$ ($1 \times 10^{-15}$)}
}
\startdata
VLBI                & $-212.5$  &	$1.1$  \\
GPS                 & $-215.7$	&   $2.0$  \\
Optical clocks      & $-213.29$	&   $0.14$   \\
\enddata
\end{deluxetable*}

\begin{deluxetable*}{lcc}
\tablecaption{Closure difference in the relative units of the
measurements obtained from the three techniques.} 
\label{tab:closure}
\tablewidth{0pt}
\tablehead{
\colhead{} & \colhead{Closure difference} & \colhead{Uncertainty} \\
\colhead{} & \colhead{$\Delta y$ ($1 \times 10^{-15}$} & \colhead{$u_\text{tot}$ ($1 \times 10^{-15}$)}
}
\startdata
VLBI  $-$  OC	    & $0.8$	   & $1.1$ \\
GPS   $-$  OC     & $-2.4$   & $2.0$ \\
VLBI  $-$  GPS	& $3.2$	   & $2.3$ \\
\enddata
\end{deluxetable*}

\section{Discussion}\label{sec:disc}

\subsection{VLBI-GPS-Optical clock comparison}

The three measurements of the average maser frequency difference obtained with VLBI, optical clocks and GPS are presented in Table~\ref{tab:hmdiff}. From these values we can calculate three closure differences as shown in Table~\ref{tab:closure}. These measurements show good agreement between the different techniques. In particular the agreement between the VLBI measurement and the optical clock measurement is within 2 sigma, confirming VLBI’s potential as a competitive intercontinental clock comparison method. We also note that the estimated frequency drift (CL2) from VLBI analysis via VieVS ($0.7(1.0)\times 10^{-15}$ day$^{-1}$) is comparable within uncertainty with the drift obtained from optical clocks ($0.7(2.7)\times 10^{-16}$ day$^{-1}$). 

Given the uncertainties achieved in this first campaign, the results are presented in terms of H-maser frequency differences. Within this framework, optical clock data are used to evaluate the VLBI link. However, these results can also be interpreted as a demonstration of VLBI-based optical clock comparison, which may serve as a valuable complement to existing intercontinental methods.

The results in Table~\ref{tab:hmdiff} confirm previous findings \citep{Pizzocaro2021} that VLBI has the potential to achieve lower uncertainties than GPS over similar durations. However, it is important to note that GPS, especially with advanced iPPP algorithms \citep{Petit2021}, can achieve uncertainties in the low $10^{-16}$ range over several days. Therefore, assessing whether standard VLBI systems can reach this level of accuracy is crucial. For evaluating this capability, continuous week-long VLBI sessions, such as geodetic campaigns, are preferable to multiple shorter campaigns. This approach, however, introduces significant challenges, including increased data volume, complex pre-processing, and more demanding analysis compared to GPS-based clock comparisons. 

\subsection{High-frequency geodetic VLBI observation}
This pilot study aimed to assess whether high-frequency VLBI observations can reliably determine clock parameters. Our K-band experiment, which employed a total bandwidth of 512 MHz composed of 16 consecutive frequency channels of 32 MHz each, achieved group delay residuals of approximately $26$ ps for both $\nu$Solve and VieVS, indicating promising performance. Further improvements are anticipated through the use of broader bandwidths and higher observing frequencies.

To enhance group delay precision, we plan to deploy Compact Triple-band Receivers (CTRs; \cite{Han2017}), capable of operating across the K/Q/W bands ($18$-$115$ GHz), thereby enabling broadband VLBI. With full frequency coverage, the expected group delay uncertainty is projected to reach the sub-picosecond level ($\sim$ 0.2 ps), as previously demonstrated by the KVN system, which operated over a frequency range of $19.584$ to $92.576$ GHz using 16 channels of $51$2 MHz each (total data rate of $32$ Gbps).

Additionally, recent results from KVN \citep{xu2024} have shown that geodetic VLBI observations at frequency up to 132 GHz are feasible, validating the approach for high-precision astrometry and geodesy. A follow-up campaign is scheduled for 2026, following the upgrade of the Medicina radio telescope with a CTR, to evaluate whether such enhancement can deliver anticipated improvements in group delay accuracy.

\subsection{Atmospheric effects: troposphere and ionosphere}

Accurate modeling of tropospheric delays is essential for reliable estimation of clock parameters. The VieVS software provided improved parameterization by employing the VMF3 mapping function and GPT3 a priori model \citep{Landskron2018}. In addition, the KVN-style simultaneous multi-frequency system has proven highly effective in calibrating tropospheric phase fluctuations in the millimeter-wave regime through the use of the Frequency Phase Transfer (FPT) method \citep{jung2011, rioja2015, rioja2020}, in which high-frequency observations are calibrated using scaled solutions from lower-frequency data. We expect that applying FPT in conjunction with CTRs and enhanced tropospheric modeling will enable effective VLBI observations at high frequencies, despite challenges such as atmospheric opacity and turbulence. This has already been demonstrated in a recent study using the KVN \citep{xu2024}. 

Ionospheric corrections applied using Global Ionospheric Maps (GIMs) in VieVS showed negligible influence on the estimated clock parameter in this experiment. This finding is consistent with results from \cite{xu2024}, who tested both GNSS-based Global Ionospheric Maps and dual-frequency VLBI combinations. Although ionospheric effects are minimal over the relatively short KVN baselines, they can become significant on longer intercontinental baselines due to global variations in Total Electronic Content (TEC), potentially introducing delays of several tens of picoseconds. 
Nonetheless, the KVN-style simultaneous multi-frequency receivers (e.g., CTRs) are expected to mitigate ionospheric impact across the 18–115 GHz range. This is because (1) the ionosphere imposes a frequency-dependent phase shift that diminishes at higher frequencies, and (2) simultaneous observations at multiple bands allow comparing and compensating dispersive delays, thereby improving phase calibration and group-delay precision—even for longer intercontinental baselines such as Korea–Italy.

A more comprehensive assessment of atmospheric effects on clock parameter estimation using high-frequency, wideband VLBI observations is planned for 2026, coinciding with the anticipated deployment of upgraded CTR systems.

\subsection{Source structure effects and instrumental delays }

\begin{figure*}[htb!]
\centering
\includegraphics[width=0.60\textwidth]{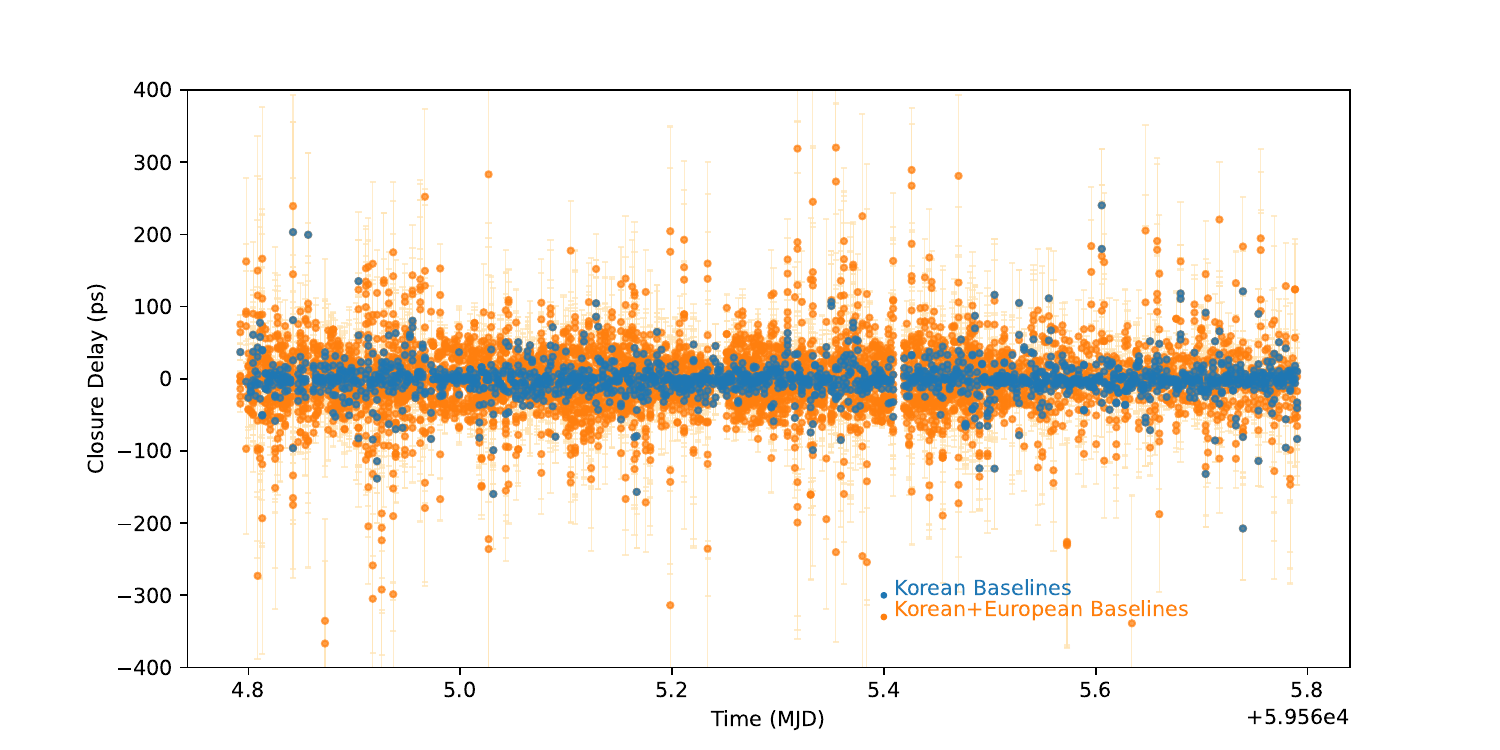}
\caption{Closure delays during the experiment. Blue and orange dots represent closure delays for Korean baselines (KVN and Sejong) and Korean-European baselines, respectively.}\label{fig:fig3}
\end{figure*}

The K-band sources used in this study were selected from the ICRF3 and AstroGeo databases. Source structure remains a major contributor to uncertainty in geodetic VLBI, particularly at lower frequencies, where non-closure delays may arise and impact baseline solutions \citep{Xu2019, Bolotin2019, Sekido2021, Pizzocaro2021}. Observations at higher frequencies, as enabled by CTRs, are expected to alleviate these issues due to the more compact nature of radio sources (e.g., \cite{charlot2010, de2023}).

Figure~\ref{fig:fig3} presents the closure delay results from our analysis, showing a WRMS of 20.1 ps across all baselines and a significantly lower value of 5.8 ps for Korean-only baselines. This discrepancy may result from: (1) source structures being resolved over longer ($\sim$ 9,000 km) baselines, (2) lower SNRs for fringe detection on those baselines, (3) the absence of instrumental phase calibration (PCAL) systems, or a combination of these and other residual errors. 

At the KVN and Sejong stations, round-trip monitoring systems are deployed to ensure stable distribution of reference frequency from the observatory building to the receiver cabin. More recently, advancements in the KVN-style multi-frequency system have introduced the integration of optical frequency comb (OFC)-based technologies. These systems generate and distribute phase-coherent, low-noise RF signals that are locked to atomic (or, optical in near future) clock references. This enables precise calibration of instrumental phase offsets and frequency-dependent variations across multiple receiver bands \citep{hyun2025}. Photonic PCAL systems of this kind are particularly  important for ultra-broadband (multi-frequency) VLBI astrometry and geodesy, including optical clock comparison experiments.

Future improvements in source structure delay modeling and the deployment of CTRs with integrated PCAL are expected to yield multiple benefits. First, incorporating more telescopes on intermediate baselines would enable more robust source imaging, aiding in the estimation and calibration of source structure effects. Second, improved instrumental calibration would allow for quantitative separation of delay contributions due to source structure versus instrumentation. This would support the identification of optimal VLBI targets (i.e., point-like sources) and enable precise evaluation of source structure effects, particularly for intercontinental baselines.

\section{Conclusion and future prospect} \label{sec:conc} 

As a pilot study for optical clock comparison using existing VLBI networks, we designed, organized, and implemented a geodetic VLBI campaign at K-band. This experiment serves as a precursor to the deployment of new high-frequency broadband receivers (e.g., CTR), which will soon be installed at several VLBI stations. This new infrastructure is expected to be utilized by many scientific research efforts, and we specifically aim to explore its application in metrology. Therefore, our pilot campaign was conducted as a standard geodetic observation, utilizing K-band, standard scheduling, and analysis techniques, with the primary goal of identifying the potential and critical points of this methodology.
The results obtained in this work with different approaches were in agreement at the $10^{-15}$ s/s level, which states that standard geodetic VLBI campaigns could be used for intercontinental clock comparisons, currently only possible via satellite techniques.

In the future, an optimal frequency setup in the range 18-115 GHz will be studied to optimize the next VLBI observations. This will exploit the new high-frequency broadband receivers (CTRs), which have demonstrated their effectiveness in improving the accuracy of the resulting group delays \citep{xu2024}. As a result, they will allow a better estimation of the clock parameters of the stations.

We plan to carry out longer experiments with a larger network including other international stations equipped with the same receivers, which will allow us to connect to the international metrology community.
This will bring with it additional challenges because of the large amount of data that will have to be stored and  transferred via the network to the correlators in Italy and Korea. Nevertheless, this network has the potential to provide the metrology community with a high-performing and independent method for intercontinental clocks comparisons.
\\
\begin{acknowledgments}
This research was supported by the National Research Council of Science and Technology (NST) grant by the Korea government (MSIT) (No. CAP22061-000) and Measurement Technology for Grand National Strategic industries funded by KRISS (KRISS-2026-GP2026-0012). The observations and correlations are supported through the high-speed network connections among the KVN and Sejong stations provided by the KREONET (Korea Research Environment Open NETwork), which is operated by the KISTI (Korea Institute of Science and Technology Information). Roberto Ricci acknowledges the support of the European Research Council through the Consolidator
grant BHianca (Grant agreement ID: 101002761). The authors thank the anonymous referee who helped improve the presentation of this work to a wider audience.
\end{acknowledgments}





%

\facilities{Medicina:32m, Sejong, KVN, Yebes:40m. }

\software{{\it SKED} \citep{Gipson2010},
          {\it DiFX}   \citep{Deller2007},
           {\it fourfit} \citep{Hoak2022},
          {\it $\nu$Solve} \citep{Bolotin2014},
         {\it VieVS} \citep{Boehm2018}}

\bibliography{occ-vlbi}{}

@article{dimarq2024,
doi = {10.1088/1681-7575/ad17d2},
url = {https://dx.doi.org/10.1088/1681-7575/ad17d2},
year = {2024},
month = {jan},
publisher = {IOP Publishing},
volume = {61},
number = {1},
pages = {012001},
author = {Dimarcq, N and Gertsvolf, M and Mileti, G and Bize, S and Oates, C W and Peik, E and Calonico, D and Ido, T and Tavella, P and Meynadier, F and Petit, G and Panfilo, G and Bartholomew, J and Defraigne, P and Donley, E A and Hedekvist, P O and Sesia, I and Wouters, M and Dubé, P and Fang, F and Levi, F and Lodewyck, J and Margolis, H S and Newell, D and Slyusarev, S and Weyers, S and Uzan, J-P and Yasuda, M and Yu, D-H and Rieck, C and Schnatz, H and Hanado, Y and Fujieda, M and Pottie, P-E and Hanssen, J and Malimon, A and Ashby, N},
title = {Roadmap towards the redefinition of the second},
journal = {Metrologia}
}

@ARTICLE{2025arXiv250118276X,
       author = {{Xu}, Ming Hui and {Charlot}, Patrick},
        title = "{Variations of absolute source positions determined from quad-band VLBI observations}",
      journal = {arXiv e-prints},
         year = 2025,
        month = jan,
          eid = {arXiv:2501.18276},
        pages = {arXiv:2501.18276},
          doi = {10.48550/arXiv.2501.18276},
archivePrefix = {arXiv},
       eprint = {2501.18276},
 primaryClass = {astro-ph.GA},
       adsurl = {https://ui.adsabs.harvard.edu/abs/2025arXiv250118276X}
}

@article{kouba,
title = {GPS precise point positioning using IGS orbit products},
journal = {Physics and Chemistry of the Earth, Part A: Solid Earth and Geodesy},
volume = {26},
number = {6},
pages = {573-578},
year = {2001},
note = {Proceedings of the First COST Action 716 Workshop Towards Operational GPS Meteorology and the Second Network Workshop of the International GPS Service (IGS)},
issn = {1464-1895},
doi = {https://doi.org/10.1016/S1464-1895(01)00103-X},
url = {https://www.sciencedirect.com/science/article/pii/S146418950100103X},
author = {P. Héroux and J. Kouba}
}

@article{CDDIS,
title = {The crustal dynamics data information system: A resource to support scientific analysis using space geodesy},
journal = {Advances in Space Research},
volume = {45},
number = {12},
pages = {1421-1440},
year = {2010},
note = {DORIS: Scientific Applications in Geodesy and Geodynamics},
issn = {0273-1177},
doi = {https://doi.org/10.1016/j.asr.2010.01.018},
url = {https://www.sciencedirect.com/science/article/pii/S0273117710000530},
author = {Carey E. Noll}
}

@article{Clivati2022topa,
  title = {Coherent Optical-Fiber Link Across Italy and France},
  author = {Clivati, C. and Pizzocaro, M. and Bertacco, E.K. and Condio, S. and Costanzo, G.A. and Donadello, S. and Goti, I. and Gozzelino, M. and Levi, F. and Mura, A. and Risaro, M. and Calonico, D. and T\o{}nnes, M. and Pointard, B. and Mazouth-Laurol, M. and Le Targat, R. and Abgrall, M. and Lours, M. and Le Goff, H. and Lorini, L. and Pottie, P.-E. and Cantin, E. and Lopez, O. and Chardonnet, C. and Amy-Klein, A.},
  journal = {Phys. Rev. Appl.},
  volume = {18},
  issue = {5},
  pages = {054009},
  numpages = {11},
  year = {2022},
  month = {Nov},
  publisher = {American Physical Society},
  doi = {10.1103/PhysRevApplied.18.054009},
  url = {https://link.aps.org/doi/10.1103/PhysRevApplied.18.054009}
}

@article{ charlot20,
author = {{Charlot}, P. and {Jacobs, C. S.} and {Gordon, D.} and {Lambert, S.} and {de Witt, A.} and {Böhm, J.} and {Fey, A. L.} and {Heinkelmann, R.} and {Skurikhina, E.} and {Titov, O.} and {Arias, E. F.} and {Bolotin, S.} and {Bourda, G.} and {Ma, C.} and {Malkin, Z.} and {Nothnagel, A.} and {Mayer, D.} and {MacMillan, D. S.} and {Nilsson, T.} and {Gaume, R.}},
title = {The third realization of the International Celestial Reference Frame by very long baseline interferometry},
DOI= "10.1051/0004-6361/202038368",
url= "https://doi.org/10.1051/0004-6361/202038368",
journal = {A and A},
year = 2020,
volume = 644,
pages = "A159",
}

@ARTICLE{xu2024,
       author = {{Xu}, Shuangjing and {Jung}, Taehyun and {Zhang}, Bo and {Xu}, Ming Hui and {Byun}, Do-Young and {He}, Xuan and {Sakai}, Nobuyuki and {Titov}, Oleg and {Shu}, Fengchun and {Kim}, Hyo-Ryoung and {Cho}, Jungho and {Yoo}, Sung-Moon and {Choi}, Byung-Kyu and {Lee}, Woo Kyoung and {Sun}, Yan and {Mai}, Xiaofeng and {Wang}, Guangli},
        title = {{A Geodetic and Astrometric VLBI Experiment at 22/43/88/132 GHz}},
      journal = {The Astronomical Journal},
        year = 2024,
        month = nov,
        volume = 168,
        issue = 5,
        doi = {10.3847/1538-3881/ad7af0},
}

@INPROCEEDINGS{Rieck2012,
  author={Rieck, Carsten and Haas, Rüdiger and Jarlemark, Per and Jaldehag, Kenneth},
  booktitle={2012 European Frequency and Time Forum}, 
  title={VLBI frequency transfer using CONT11}, 
  year={2012},
  volume={},
  number={},
  pages={163-165},
  doi={10.1109/EFTF.2012.6502358}}

@ARTICLE{Counselman1977,
  author={Counselman, C.C. and Shapiro, I.I. and Rogers, A.E.E. and Hinteregger, H.F. and Knight, C.A. and Whitney, A.R. and Clark, T.A.},
  journal={Proceedings of the IEEE}, 
  title={VLBI clock synchronization}, 
  year={1977},
  volume={65},
  number={11},
  pages={1622-1623},
  doi={10.1109/PROC.1977.10793}}

@Article{Bauch2005a,
  Title                    = {Comparison between frequency standards in {Europe} and the {USA} at the 10-15 uncertainty level},
  Author                   = {A Bauch and J Achkar and S Bize and D Calonico and R Dach and R Hlava{\'{c}} and L Lorini and T Parker and G Petit and D Piester and K Szymaniec and P Uhrich},
  Journal                  = {Metrologia},
  Year                     = {2005},

  Month                    = {dec},
  Number                   = {1},
  Pages                    = {109--120},
  Volume                   = {43},
  Doi                      = {10.1088/0026-1394/43/1/016},
  Publisher                = {{IOP} Publishing},
  Url                      = {https://doi.org/10.1088%2F0026-1394%2F43%2F1%2F016}
}

@InProceedings{Bolotin2014,
  Title                    = {The VLBI data analysis software ?solve: development progress and plans for the future},
  Author                   = {Bolotin, S. and Baver, K. and Gipson, J. and Gordon, D. and MacMillan, D.},
  Booktitle                = {IVS 2014 general meeting proceedings VGOS: the new VLBI network},
  Year                     = {2014}
}

@InProceedings{Bolotin2019,
  Title                    = {The source structure effect in broadband observations},
  Author                   = {Bolotin, S. and Baver, K. and Bolotina, O. and Gipson, J. and Gordon, D. and Le Bail, K. and MacMillan, D.},
  Booktitle                = {Proc. of 24th Meeting of the European VLBI Group for Geodesy and Astrometry},
  Year                     = {2019}
}

@Article{Brewer2019,
  Title                    = {{$^{27}{\mathrm{Al}}^{+}$} Quantum-Logic Clock with a Systematic Uncertainty below ${10}^{\ensuremath{-}18}$},
  Author                   = {Brewer, S. M. and Chen, J.-S. and Hankin, A. M. and Clements, E. R. and Chou, C. W. and Wineland, D. J. and Hume, D. B. and Leibrandt, D. R.},
  Journal                  = {Physical Review Letters},
  Year                     = {2019},

  Month                    = {Jul},
  Pages                    = {033201},
  Volume                   = {123},

  Doi                      = {10.1103/PhysRevLett.123.033201},
  Issue                    = {3},
  Numpages                 = {6},
  Publisher                = {American Physical Society},
  Url                      = {https://link.aps.org/doi/10.1103/PhysRevLett.123.033201}
}

@misc{ledbetter25,
  Title                    = {Iodine Optical Clock use at the Event Horizon Telescope for VLBI},
  Author                   = {Ledbetter, M. and Kowligy, A. and Roslund, J. and Cingoz, A. and Partridge, G. and Patel, P. and Pashollari, E. and Popp, E. and Roller, F. and Sheredy, D. and Skulason, G. and Song, J. and Atchison, E. and Husain, O. and Carney, P. and Pasha, M.K. and Rakholia, A. and Dowd, A. and Abo-Shaeer, J. and Boyd, M. and Marrone, D. and Reiland G.},
  Booktitle                = {Presentiation at the 2025 International Technical Meeting of The Institute of Navigation},
  Year                     = {2025},
  Url                      = {https://www.ion.org/itm/abstracts.cfm?paperID=15127}
}

@Article{Chiodo2015,
  author    = {Chiodo, Nicola and Quintin, Nicolas and Stefani, Fabio and Wiotte, Fabrice and Camisard, Emilie and Chardonnet, Christian and Santarelli, Giorgio and Amy-Klein, Anne and Pottie, Paul-Eric and Lopez, Olivier},
  title     = {Cascaded optical fiber link using the internet network for remote clocks comparison},
  journal   = {Opt. Express},
  year      = {2015},
  volume    = {23},
  number    = {26},
  pages     = {33927--33937},
  month     = dec,
  publisher = {OSA},
  url       = {http://opg.optica.org/oe/abstract.cfm?URI=oe-23-26-33927},
}

@Article{Clark1979,
  Title                    = {Optical isotopic shifts and hyperfine splittings for {Yb}},
  Author                   = {Clark, D. L. and Cage, M. E. and Lewis, D. A. and Greenlees, G. W.},
  Journal                  = {Phys. Rev. A},
  Year                     = {1979},

  Month                    = {Jul},
  Pages                    = {239--253},
  Volume                   = {20},

  Doi                      = {10.1103/PhysRevA.20.239},
  Issue                    = {1},
  Publisher                = {American Physical Society},
  Url                      = {http://link.aps.org/doi/10.1103/PhysRevA.20.239}
}

@InProceedings{Serrano2009,
  Title                    = {The White Rabbit Project},
  Author                   = {J. Serrano and P. Alvarez and M. Cattin and others},
  Booktitle                = {Int. Conf. Accelerator and Large Experimental Physics Control Systems (ICALEPCS), Kobe, Japan, 2009.},
  Year                     = {2009},
  Url                      = {}
}

@Article{Clivati2015,
  Title                    = {A coherent fiber link for very long baseline interferometry},
  Author                   = {C. Clivati and G. A. Costanzo and M. Frittelli and F. Levi and A. Mura and M. Zucco and R. Ambrosini and C. Bortolotti and F. Perini and M. Roma and D. Calonico},
  Journal                  = {IEEE Transactions on Ultrasonics, Ferroelectrics, and Frequency Control},
  Year                     = {2015},

  Month                    = {November},
  Number                   = {11},
  Pages                    = {1907-1912},
  Volume                   = {62},

  Doi                      = {10.1109/TUFFC.2015.007221},
  ISSN                     = {0885-3010}
}

@Article{Clivati2020,
  author    = {Cecilia Clivati and Paolo Savio and Silvio Abrate and Vittorio Curri and Roberto Gaudino and Marco Pizzocaro and Davide Calonico},
  title     = {Robust optical frequency dissemination with a dual-polarization coherent receiver},
  journal   = {Opt. Express},
  year      = {2020},
  volume    = {28},
  number    = {6},
  month     = {Mar},
  pages     = {8494--8511},
  doi       = {10.1364/OE.378602},
  url       = {http://www.opticsexpress.org/abstract.cfm?URI=oe-28-6-8494},
  publisher = {OSA},
}

@Article{Dierikx2016,
  Title                    = {White Rabbit Precision Time Protocol on Long-Distance Fiber Links},
  Author                   = {E. F. Dierikx and A. E. Wallin and T. Fordell and J. Myyry and P. Koponen and M. Merimaa and T. J. Pinkert and J. C. J. Koelemeij and H. Z. Peek and R. Smets},
  Journal                  = {IEEE Transactions on Ultrasonics, Ferroelectrics, and Frequency Control},
  Year                     = {2016},

  Month                    = {July},
  Number                   = {7},
  Pages                    = {945-952},
  Volume                   = {63},

  Doi                      = {10.1109/TUFFC.2016.2518122},
  ISSN                     = {0885-3010}
}

@Article{Fujieda2014,
  Title                    = {Carrier-phase two-way satellite frequency transfer over a very long baseline},
  Author                   = {M Fujieda and D Piester and T Gotoh and J Becker and M Aida and A Bauch},
  Journal                  = {Metrologia},
  Year                     = {2014},

  Month                    = {may},
  Number                   = {3},
  Pages                    = {253--262},
  Volume                   = {51},
  Doi                      = {10.1088/0026-1394/51/3/253},
  Publisher                = {{IOP} Publishing},
  Url                      = {https://doi.org/10.1088%2F0026-1394%2F51%2F3%2F253}
}

@Article{Gill2016,
  Title                    = {Is the time right for a redefinition of the second by optical atomic clocks?},
  Author                   = {Patrick Gill},
  Journal                  = {Journal of Physics: Conference Series},
  Year                     = {2016},
  Number                   = {1},
  Pages                    = {012053},
  Volume                   = {723},
  Url                      = {http://stacks.iop.org/1742-6596/723/i=1/a=012053}
}

@InProceedings{Gipson2010,
  Title                    = {An Introduction to Sked},
  Author                   = {John Gipson},
  Booktitle                = {IVS 2010 General Meeting Proceedings},
  Year                     = {2010},
  Series                   = {NASA/CP 2010-215864},
  Pages                    = {77--84},
 }

@InProceedings{Gordon2017,
  Title                    = {{GSFC VLBI Analysis Center Report}},
  Author                   = {Gordon, David and Ma, Chopo and MacMillan, Dan and Gipson, John and Bolotin, Sergei and Le Bail, Karine and Baver, Karen},
  Booktitle                = {International VLBI Service for Geodesy and Astrometry 2015+2016 Biennial Report},
  Year                     = {2017},
  Editor                   = {Baver, K. D. and Behrend, D. and Armstrong, Kyla L.},
  Month                    = {dec},
  Series                   = {NASA/TP–2017–219021},
  Url                      = {http://ivscc.gsfc.nasa.gov/publications/br_2015_2016/index.html}
}

@Article{Grebing2016,
  Title                    = {Realization of a timescale with an accurate optical lattice clock},
  Author                   = {Christian Grebing and Ali Al-Masoudi and S\"{o}ren D\"{o}rscher and Sebastian H\"{a}fner and Vladislav Gerginov and Stefan Weyers and Burghard Lipphardt and Fritz Riehle and Uwe Sterr and Christian Lisdat},
  Journal                  = {Optica},
  Year                     = {2016},

  Month                    = {Jun},
  Number                   = {6},
  Pages                    = {563--569},
  Volume                   = {3},
  Doi                      = {10.1364/OPTICA.3.000563},
  Publisher                = {OSA},
  Url                      = {http://www.osapublishing.org/optica/abstract.cfm?URI=optica-3-6-563}
}

@Article{Grotti2018,
  Title                    = {Geodesy and metrology with a transportable optical clock},
  Author                   = {Grotti, Jacopo and Koller, Silvio and Vogt, Stefan and Häfner, Sebastian and Sterr, Uwe and Lisdat, Christian and Denker, Heiner and Voigt, Christian and Timmen, Ludger and Rolland, Antoine and Baynes, Fred N. and Margolis, Helen S. and Zampaolo, Michel and Thoumany, Pierre and Pizzocaro, Marco and Rauf, Benjamin and Bregolin, Filippo and Tampellini, Anna and Barbieri, Piero and Zucco, Massimo and Costanzo, Giovanni A. and Clivati, Cecilia and Levi, Filippo and Calonico, Davide},
  Journal                  = {Nature Physics},
  Year                     = {2018},
  Number                   = {5},
  Pages                    = {437--441},
  Volume                   = {14},
  ISSN                     = {1745-2481},
  Refid                    = {Grotti2018},
  Url                      = {https://doi.org/10.1038/s41567-017-0042-3}
}

@Article{Hachisu2014,
  Title                    = {Direct comparison of optical lattice clocks with an intercontinental baseline of 9000 km},
  Author                   = {H. Hachisu and M. Fujieda and S. Nagano and T. Gotoh and A. Nogami and T. Ido and St. Falke and N. Huntemann and C. Grebing and B. Lipphardt and Ch. Lisdat and D. Piester},
  Journal                  = {Optics Letters},
  Year                     = {2014},

  Month                    = {Jul},
  Number                   = {14},
  Pages                    = {4072--4075},
  Volume                   = {39},
  Doi                      = {10.1364/OL.39.004072},
  Publisher                = {OSA},
  Url                      = {http://ol.osa.org/abstract.cfm?URI=ol-39-14-4072}
}

@Article{Hachisu2015,
  Title                    = {Intermittent optical frequency measurements to reduce the dead time uncertainty of frequency link},
  Author                   = {Hidekazu Hachisu and Tetsuya Ido},
  Journal                  = {Japanese Journal of Applied Physics},
  Year                     = {2015},
  Number                   = {11},
  Pages                    = {112401},
  Volume                   = {54},
  Url                      = {http://stacks.iop.org/1347-4065/54/i=11/a=112401}
}

@Article{Hachisu2017,
  Title                    = {{SI-traceable measurement of an optical frequency at the low \num{1e-16} level without a local primary standard}},
  Author                   = {Hidekazu Hachisu and G\'{e}rard Petit and Fumimaru Nakagawa and Yuko Hanado and Tetsuya Ido},
  Journal                  = {Optics Express},
  Year                     = {2017},

  Month                    = {Apr},
  Number                   = {8},
  Pages                    = {8511--8523},
  Volume                   = {25},
  Doi                      = {10.1364/OE.25.008511},
  Publisher                = {OSA},
  Url                      = {http://www.opticsexpress.org/abstract.cfm?URI=oe-25-8-8511}
}

@article{Han2017,
	affiliation = {Korea Astronomy and Space Science Institute; Jet Propulsion Laboratory},
	author = {Han, Seog-Tae and Lee, Jung-Won and Lee, Bangwon and Chung, Moon-Hee and Lee, Sung-Mo and Je, Do-Heung and Wi, Seog-Oh and Goldsmith, Paul F.},
	copyright = {Springer Science+Business Media, LLC},
	doi = {10.1007/s10762-017-0438-2},
	journal = {Journal of Infrared, Millimeter, and Terahertz Waves},
	language = {English},
	month = {9},
	number = {12},
	pages = {1487-1501},
	title = {A Millimeter-Wave Quasi-Optical Circuit for Compact Triple-Band Receiving System},
	volume = {38},
	year = {2017},
}

@techreport{Hurd1979,
    author = {W. J. Hurd and S. C. Wardrip and J. Bussion and J. Oaks and T. McCaskill and H. Warren and G. Whitworth},
    title = {Submicrosecond Comparison of Intercontinental Clock\r\nSynchronization by VLBI and the NTS Satellite. The Deep Space Network Progress Report},
    institution = {Jet Propulsion Laboratory},
    year = {1979},
    Volume = {42-29},
    Url =  {https://ipnpr.jpl.nasa.gov/progress_report/42-49/49J.PDF}
}

@article{jung2011,
    author = {Jung, Taehyun and Sohn, Bong Won and Kobayashi, Hideyuki and Sasao, Tetsuo and Hirota, Tomoya and Kameya, Osamu and Choi, Yoon Kyung and Chung, Hyun Soo},
    title = {First Simultaneous Dual-Frequency Phase Referencing VLBI Observation with VERA},
    journal = {Publications of the Astronomical Society of Japan},
    volume = {63},
    number = {2},
    pages = {375-385},
    year = {2011},
    month = {04},
    issn = {0004-6264},
    doi = {10.1093/pasj/63.2.375},
}

@article{krehlik17,
	author = {{Krehlik, P.} and {Buczek, L.} and {Kolodziej, J.} and {Lipi\'{}nski, M.} and {\'{}Sliwczy\'{}nski, L.} and {Nawrocki, J.} and {Noga\'{}s, P.} and {Marecki, A.} and {Pazderski, E.} and {Ablewski, P.} and {Bober, M.} and {Ciurylo, R.} and {Cygan, A.} and {Lisak, D.} and {Maslowski, P.} and {Morzy\'{}nski, P.} and {Zawada, M.} and {Campbell, R. M.} and {Pieczerak, J.} and {Binczewski, A.} and {Turza, K.}},
	title = {Fibre-optic delivery of time and frequency to VLBI station},
	DOI= "10.1051/0004-6361/201730615",
	url= "https://doi.org/10.1051/0004-6361/201730615",
	journal = {A\&A},
	year = 2017,
	volume = 603,
	pages = "A48",
}

@Article{Landskron2018,
  Title                    = {{VMF3/GPT3: refined discrete and empirical troposphere mapping functions}},
  Author                   = { Landskron, Daniel and B{\"o}hm, Johannes},
  Journal                  = {Journal of Geodesy},
  Year                     = {2018},

  Month                    = {Apr},
  Number                   = {4},
  Pages                    = {349--360},
  Volume                   = {92},
  Day                      = {01},
  Doi                      = {10.1007/s00190-017-1066-2},
  ISSN                     = {1432-1394},
  Url                      = {https://doi.org/10.1007/s00190-017-1066-2}
}

@Article{Leute2016,
  Title                    = {Frequency Comparison of {$^{171}{\text {Yb}}^+$} Ion Optical Clocks at {PTB} and {NPL} via {GPS PPP}},
  Author                   = {J. Leute and N. Huntemann and B. Lipphardt and C. Tamm and P. B. R. Nisbet-Jones and S. A. King and R. M. Godun and J. M. Jones and H. S. Margolis and P. B. Whibberley and A. Wallin and M. Merimaa and P. Gill and E. Peik},
  Journal                  = {IEEE Transactions on Ultrasonics, Ferroelectrics, and Frequency Control},
  Year                     = {2016},

  Month                    = {July},
  Number                   = {7},
  Pages                    = {981-985},
  Volume                   = {63},

  Doi                      = {10.1109/TUFFC.2016.2524988},
  ISSN                     = {0885-3010}
}

@Article{Lisdat2016,
  Title                    = {A clock network for geodesy and fundamental science},
  Author                   = {Lisdat, C. and Grosche, G. and Quintin, N. and Shi, C. and Raupach, S.M.F. and Grebing, C. and Nicolodi, D. and Stefani, F. and Al-Masoudi, A. and D\"orscher, S. and H\"afner, S. and Robyr, J.-L. and Chiodo, N. and Bilicki, S. and Bookjans, E. and Koczwara, A. and Koke, S. and Kuhl, A. and Wiotte, F. and Meynadier, F. and Camisard, E. and Abgrall, M. and Lours, M. and Legero, T. and Schnatz, H. and Sterr, U. and Denker, H. and Chardonnet, C. and Le Coq, Y. and Santarelli, G. and Amy-Klein, A. and Le Targat, R. and Lodewyck, J. and Lopez, O and Pottie, P.-E.},
  Journal                  = {Nature Communications},
  Year                     = {2016},

  Month                    = aug,
  Pages                    = {12443},
  Volume                   = {7},
  Publisher                = {The Author(s)},
  Url                      = {http://dx.doi.org/10.1038/ncomms12443}
}

@Article{McGrew2018,
  Title                    = {Atomic clock performance enabling geodesy below the centimetre level},
  Author                   = {McGrew, W. F. and Zhang, X. and Fasano, R. J. and Schäffer, S. A. and Beloy, K. and Nicolodi, D. and Brown, R. C. and Hinkley, N. and Milani, G. and Schioppo, M. and Yoon, T. H. and Ludlow, A. D.},
  Journal                  = {Nature},
  Year                     = {2018},
  Number                   = {7734},
  Pages                    = {87--90},
  Volume                   = {564},
  ISSN                     = {1476-4687},
  Refid                    = {McGrew2018},
  Url                      = {https://doi.org/10.1038/s41586-018-0738-2}
}

@Article{Musha2008,
  author        = {Musha, Mitsuru and Hong, Feng-Lei and Nakagawa, Ken’ichi and Ueda, Ken-ichi},
  title         = {Coherent optical frequency transfer over 50-km physical distance using a 120-km-long installed telecom fiber network},
  journal       = {Opt. Express},
  year          = {2008},
  volume        = {16},
  number        = {21},
  month         = oct,
  pages         = {16459--16466},
  url           = {http://opg.optica.org/oe/abstract.cfm?URI=oe-16-21-16459},
  publisher     = {OSA},
}

@Article{Nakamura2020,
  Title                    = {Coherent optical clock down-conversion for microwave frequencies with 10-18 instability},
  Author                   = {Nakamura, Takuma and Davila-Rodriguez, Josue and Leopardi, Holly and Sherman, Jeff A. and Fortier, Tara M. and Xie, Xiaojun and Campbell, Joe C. and McGrew, William F. and Zhang, Xiaogang and Hassan, Youssef S. and Nicolodi, Daniele and Beloy, Kyle and Ludlow, Andrew D. and Diddams, Scott A. and Quinlan, Franklyn},
  Journal                  = {Science},
  Year                     = {2020},
  Number                   = {6493},
  Pages                    = {889--892},
  Volume                   = {368},
  Doi                      = {10.1126/science.abb2473},
  ISSN                     = {0036-8075},
  Publisher                = {American Association for the Advancement of Science},
  Url                      = {https://science.sciencemag.org/content/368/6493/889}
}

@InProceedings{Oh2010,
  Title                    = {Round-trip System Available to Measure Path Length Variation in Korea VLBI System for Geodesy},
  Author                   = {Hongjong Oh and Tetsuro Kondo and Jinoo Lee and Tuhwan Kim and Myungho Kim and Suchul Kim and Jinsik Park and Hyunhee Ju},
  Booktitle                = {IVS 2010 General Meeting Proceedings},
  Year                     = {2010},
  Series                   = {NASA/CP 2010-215864},
  Pages                    = {449--453},
 }

@Article{Pizzocaro2020,
  Title                    = {Absolute frequency measurement of the \ce{^1S_0} -- \ce{^3P_0} transition~of~\ce{^{171}Yb} with a link to international atomic time},
  Author                   = {Marco Pizzocaro and Filippo Bregolin and Piero Barbieri and Benjamin Rauf and Filippo Levi and Davide Calonico},
  Journal                  = {Metrologia},
  Year                     = {2020},

  Month                    = {may},
  Number                   = {3},
  Pages                    = {035007},
  Volume                   = {57},
  Doi                      = {10.1088/1681-7575/ab50e8},
  Publisher                = {{IOP} Publishing},
  Url                      = {https://doi.org/10.1088%2F1681-7575%2Fab50e8}
}

@Article{Riedel2020,
  Title                    = {Direct comparisons of {European} primary and secondary frequency standards via satellite techniques},
  Author                   = {Franziska Riedel and Ali Al-Masoudi and Erik Benkler and Sören Dörscher and Vladislav Gerginov and Christian Grebing and Sebastian Häfner and Nils Huntemann and Burghard Lipphardt and Christian Lisdat and Ekkehard Peik and Dirk Piester and Christian Sanner and Christian Tamm and Stefan Weyers and H Denker and Ludger Timmen and Christian Voigt and Davide Calonico and Giancarlo Cerretto and Giovanni A Costanzo and Filippo Levi and Ilaria Sesia and Joseph Achkar and Jocelyne Guéna and Michel Abgrall and Giovanni Daniele Rovera and Baptiste Chupin and Chunyan Shi and S?awomir Bilicki and Eva Bookjans and Jérôme Lodewyck and Rodolphe Le Targat and Pacome Delva and Sebastien Bize and Fred N Baynes and Charles Baynham and William Bowden and Patrick Gill and R M Godun and Ian Robert Hill and Richard Hobson and J M Jones and Steven A King and Peter Nisbet-Jones and A Rolland and S L Shemar and Peter B Whibberley and Helen S Margolis},
  Journal                  = {Metrologia},
  Year                     = {2020},
  Note                     = {not yet published in print},
  Url                      = {https://doi.org/10.1088/1681-7575/ab6745}
}

@Article{Riehle2018,
  Title                    = {The {CIPM} list of recommended frequency standard values: guidelines and procedures},
  Author                   = {Fritz Riehle and Patrick Gill and Felicitas Arias and Lennart Robertsson},
  Journal                  = {Metrologia},
  Year                     = {2018},
  Number                   = {2},
  Pages                    = {188--200},
  Volume                   = {55},
  Url                      = {http://stacks.iop.org/0026-1394/55/i=2/a=188}
}

@Article{Sanner2019,
  Title                    = {Optical clock comparison for {Lorentz} symmetry testing},
  Author                   = {Sanner, Christian and Huntemann, Nils and Lange, Richard and Tamm, Christian and Peik, Ekkehard and Safronova, Marianna S. and Porsev, Sergey G.},
  Journal                  = {Nature},
  Year                     = {2019},
  Number                   = {7747},
  Pages                    = {204--208},
  Volume                   = {567},
  ISSN                     = {1476-4687},
  Refid                    = {Sanner2019},
  Url                      = {https://doi.org/10.1038/s41586-019-0972-2}
}

@Article{Schuh2012,
  Title                    = {{VLBI:} A fascinating technique for geodesy and astrometry},
  Author                   = {H. Schuh and D. Behrend},
  Journal                  = {Journal of Geodynamics},
  Year                     = {2012},
  Pages                    = {68--80},
  Volume                   = {61},
  Doi                      = {https://doi.org/10.1016/j.jog.2012.07.007},
  ISSN                     = {0264-3707},
  Url                      = {http://www.sciencedirect.com/science/article/pii/S0264370712001159}
}

@Article{Sekido2021,
  author    = {Sekido, Mamoru and Takefuji, Kazuhiro and Ujihara, Hideki and Kondo, Tetsuro and Tsutsumi, Masanori and Kawai, Eiji and Hachisu, Hidekazu and Nemitz, Nils and Pizzocaro, Marco and Clivati, Cecilia and Perini, Federico and Negusini, Monia and Maccaferri, Giuseppe and Ricci, Roberto and Roma, Mauro and Bortolotti, Claudio and Namba, Kunitaka and Komuro, Jun’ichi and Ichikawa, Ryuichi and Suzuyama, Tomonari and Watabe, Ken-ichi and Leute, Julia and Petit, Gérard and Calonico, Davide and Ido, Tetsuya},
  title     = {A broadband {VLBI} system using transportable stations for geodesy and metrology: an alternative approach to the {VGOS} concept},
  journal   = {Journal of Geodesy},
  year      = {2021},
  volume    = {95},
  number    = {4},
  month     = mar,
  pages     = {41},
  issn      = {1432-1394},
  url       = {https://doi.org/10.1007/s00190-021-01479-8},
  refid     = {Sekido2021},
}

@Article{Takamoto2020,
  Title                    = {Test of general relativity by a pair of transportable optical lattice clocks},
  Author                   = {Takamoto, Masao and Ushijima, Ichiro and Ohmae, Noriaki and Yahagi, Toshihiro and Kokado, Kensuke and Shinkai, Hisaaki and Katori, Hidetoshi},
  Journal                  = {Nature Photonics},
  Year                     = {2020},

  Month                    = apr,
  Pages                    = {411--415},
  ISSN                     = {1749-4893},
  Refid                    = {Takamoto2020},
  Url                      = {https://doi.org/10.1038/s41566-020-0619-8}
}

@Article{Ushijima2015,
  Title                    = {Cryogenic optical lattice clocks},
  Author                   = {Ushijima, Ichiro and Takamoto, Masao and Das, Manoj and Ohkubo, Takuya and Katori, Hidetoshi},
  Journal                  = {Nature Photonics},
  Year                     = {2015},

  Month                    = mar,
  Number                   = {3},
  Pages                    = {185--189},
  Volume                   = {9},
  ISSN                     = {1749-4885},
  Publisher                = {Nature Publishing Group},
  Url                      = {http://dx.doi.org/10.1038/nphoton.2015.5}
}

@Article{Xu2019,
  Title                    = {Structure Effects for 3417 Celestial Reference Frame Radio Sources},
  Author                   = {M. H. Xu and J. M. Anderson and R. Heinkelmann and S. Lunz and H. Schuh and G. L. Wang},
  Journal                  = {The Astrophysical Journal Supplement Series},
  Year                     = {2019},

  Month                    = {may},
  Number                   = {1},
  Pages                    = {5},
  Volume                   = {242},

  Doi                      = {10.3847/1538-4365/ab16ea},
  Publisher                = {American Astronomical Society},
  Url                      = {https://doi.org/10.3847%2F1538-4365%2Fab16ea}
}

@Article{Yu2007,
  Title                    = {Uncertainty of a frequency comparison with distributed dead time and measurement interval offset},
  Author                   = {Dai-Hyuk Yu and Marc Weiss and Thomas E Parker},
  Journal                  = {Metrologia},
  Year                     = {2007},
  Number                   = {1},
  Pages                    = {91--96},
  Volume                   = {44},
  Url                      = {http://stacks.iop.org/0026-1394/44/i=1/a=014}
}

@article{Margolis2024,
doi = {10.1088/1681-7575/ad3afc},
url = {https://dx.doi.org/10.1088/1681-7575/ad3afc},
year = {2024},
month = {apr},
publisher = {IOP Publishing},
volume = {61},
number = {3},
pages = {035005},
author = {H S Margolis and G Panfilo and G Petit and C Oates and T Ido and S Bize},
title = {The CIPM list ‘Recommended values of standard frequencies’: 2021 update},
journal = {Metrologia}
}

@Article{Lodewyck2019,
  author    = {J{\'{e}}r{\^{o}}me Lodewyck},
  title     = {On a definition of the {SI} second with a set of optical clock transitions},
  journal   = {Metrologia},
  year      = {2019},
  volume    = {56},
  number    = {5},
  month     = {sep},
  pages     = {055009},
  doi       = {10.1088/1681-7575/ab3a82},
  url       = {https://doi.org/10.1088%2F1681-7575%2Fab3a82},
  publisher = {{IOP} Publishing},
}

@Article{Pizzocaro2021,
  author    = {Pizzocaro, Marco and Sekido, Mamoru and Takefuji, Kazuhiro and Ujihara, Hideki and Hachisu, Hidekazu and Nemitz, Nils and Tsutsumi, Masanori and Kondo, Tetsuro and Kawai, Eiji and Ichikawa, Ryuichi and Namba, Kunitaka and Okamoto, Yoshihiro and Takahashi, Rumi and Komuro, Junichi and Clivati, Cecilia and Bregolin, Filippo and Barbieri, Piero and Mura, Alberto and Cantoni, Elena and Cerretto, Giancarlo and Levi, Filippo and Maccaferri, Giuseppe and Roma, Mauro and Bortolotti, Claudio and Negusini, Monia and Ricci, Roberto and Zacchiroli, Giampaolo and Roda, Juri and Leute, Julia and Petit, Gérard and Perini, Federico and Calonico, Davide and Ido, Tetsuya},
  title     = {Intercontinental comparison of optical atomic clocks through very long baseline interferometry},
  journal   = {Nature Physics},
  year      = {2021},
  volume    = {17},
  number    = {2},
  month     = feb,
  pages     = {223--227},
  issn      = {1745-2481},
  url       = {https://doi.org/10.1038/s41567-020-01038-6},
  refid     = {Pizzocaro2021},
}

@Article{Nemitz2021,
  author    = {Nemitz, Nils and Gotoh, Tadahiro and Nakagawa, Fumimaru and Ito, Hiroyuki and Hanado, Yuko and Ido, Tetsuya and Hachisu, Hidekazu},
  title     = {Absolute frequency of \ce{^{87}Sr} at \num{1.8e-16} uncertainty by reference to remote primary frequency standards},
  journal   = {Metrologia},
  year      = {2021},
  volume    = {58},
  number    = {2},
  month     = feb,
  pages     = {025006},
  issn      = {1681-7575},
  url       = {http://dx.doi.org/10.1088/1681-7575/abc232},
  publisher = {IOP Publishing},
}

@Article{Beloy2021,
  author    = {Beloy, Kyle and Bodine, Martha I. and Bothwell, Tobias and Brewer, Samuel M. and Bromley, Sarah L. and Chen, Jwo-Sy and {Boulder Atomic Clock Optical Network (BACON) Collaboration} and others },
  title     = {Frequency ratio measurements at 18-digit accuracy using an optical clock network},
  journal   = {Nature},
  year      = {2021},
  volume    = {591},
  number    = {7851},
  month     = mar,
  pages     = {564--569},
  issn      = {1476-4687},
  url       = {https://doi.org/10.1038/s41586-021-03253-4},
  refid     = {Beloy2021},
}

@Article{Lange2021,
  author    = {Lange, R. and Huntemann, N. and Rahm, J. M. and Sanner, C. and Shao, H. and Lipphardt, B. and Tamm, Chr. and Weyers, S. and Peik, E.},
  title     = {Improved Limits for Violations of Local Position Invariance from Atomic Clock Comparisons},
  journal   = {Phys. Rev. Lett.},
  year      = {2021},
  volume    = {126},
  number    = {1},
  month     = jan,
  pages     = {011102},
  url       = {https://link.aps.org/doi/10.1103/PhysRevLett.126.011102},
  publisher = {American Physical Society},
  refid     = {10.1103/PhysRevLett.126.011102},
}

@Article{Kim2021,
  author    = {Kim, Huidong and Heo, Myoung-Sun and Park, Chang Yong and Yu, Dai-Hyuk and Lee, Won-Kyu},
  title     = {Absolute frequency measurement of the \ce{^{171}Yb} optical lattice clock at {KRISS} using {TAI} for over a year},
  journal   = {Metrologia},
  year      = {2021},
  volume    = {58},
  number    = {5},
  month     = aug,
  pages     = {055007},
  issn      = {1681-7575},
  url       = {http://dx.doi.org/10.1088/1681-7575/ac1950},
  publisher = {IOP Publishing},
}

@Article{Akatsuka2020,
  author    = {Akatsuka, Tomoya and Goh, Takashi and Imai, Hiromitsu and Oguri, Katsuya and Ishizawa, Atsushi and Ushijima, Ichiro and Ohmae, Noriaki and Takamoto, Masao and Katori, Hidetoshi and Hashimoto, Toshikazu and Gotoh, Hideki and Sogawa, Tetsuomi},
  title     = {Optical frequency distribution using laser repeater stations with planar lightwave circuits},
  journal   = {Opt. Express},
  year      = {2020},
  volume    = {28},
  number    = {7},
  month     = mar,
  pages     = {9186--9197},
  url       = {http://opg.optica.org/oe/abstract.cfm?URI=oe-28-7-9186},
  publisher = {OSA},
}

@Article{Husmann2021,
  author    = {Husmann, Dominik and Bernier, Laurent-Guy and Bertrand, Mathieu and Calonico, Davide and Chaloulos, Konstantinos and Clausen, Gloria and Clivati, Cecilia and Faist, Jérôme and Heiri, Ernst and Hollenstein, Urs and Johnson, Anatoly and Mauchle, Fabian and Meir, Ziv and Merkt, Frédéric and Mura, Alberto and Scalari, Giacomo and Scheidegger, Simon and Schmutz, Hansjürg and Sinhal, Mudit and Willitsch, Stefan and Morel, Jacques},
  title     = {SI-traceable frequency dissemination at 1572.06 nm in a stabilized fiber network with ring topology},
  journal   = {Opt. Express},
  year      = {2021},
  volume    = {29},
  number    = {16},
  month     = aug,
  pages     = {24592--24605},
  url       = {http://opg.optica.org/oe/abstract.cfm?URI=oe-29-16-24592},
  publisher = {OSA},
}

@Article{Droste2013,
  author    = {Droste, S. and Ozimek, F. and Udem, Th. and Predehl, K. and Hänsch, T. W. and Schnatz, H. and Grosche, G. and Holzwarth, R.},
  title     = {Optical-Frequency Transfer over a Single-Span 1840 km Fiber Link},
  journal   = {Phys. Rev. Lett.},
  year      = {2013},
  volume    = {111},
  number    = {11},
  month     = sep,
  pages     = {110801},
  url       = {https://link.aps.org/doi/10.1103/PhysRevLett.111.110801},
  publisher = {American Physical Society},
  refid     = {10.1103/PhysRevLett.111.110801},
}

@Article{Petit2021,
  author        = {Petit, Gérard},
  title         = {Sub-$10^{-16}$ accuracy {GNSS} frequency transfer with {IPPP}},
  journal       = {GPS Solutions},
  year          = {2021},
  volume        = {25},
  number        = {1},
  month         = jan,
  pages         = {22},
  issn          = {1521-1886},
  url           = {https://doi.org/10.1007/s10291-020-01062-2},
  __markedentry = {[marco:6]},
  abstract      = {Precise point positioning (PPP) using dual-frequency GNSS code and phase measurements has been the technique of choice for time and frequency transfer for more than a decade. Several analysis centers providing the satellite orbit and clock products used in PPP now make use of ambiguity resolution methods that preserve the integer nature of the carrier phase ambiguities, thus allowing the development of integer ambiguity PPP techniques (IPPP). By comparison with independent highly stable time transfer techniques not GNSS based, we show that IPPP provides time transfer stability and frequency transfer accuracy with improved long-term performance of order 7?×?10-16/T, where T is the duration in days of continuous phase measurements, thus reaching the sub???10-16 level after one week of averaging. This method can readily be put into operation between any two locations equipped with such GNSS receivers which have become ubiquitous equipment in scientific institutes. The analysis procedure is described in some detail, summarizing lessons learned from many years of experience and pointing out the necessary checks and the practical points that are needed to ensure that optimal results will be obtained.},
  file          = {Petit2021.pdf:Articles/Petit2021.pdf:PDF},
  refid         = {Petit2021},
  timestamp     = {2022-04-05},
}

@INPROCEEDINGS{Bolotin2017,
       author = {{Bolotin}, S. and {Baver}, K. and {Gipson}, J. and {Gordon}, D. and {MacMillan}, D.},
        title = "{Implementation of the vgosDb Format at the GSFC VLBI Analysis Center}",
    booktitle = {23rd European VLBI Group for Geodesy and Astrometry Working Meeting},
         year = 2017,
       editor = {{Haas}, R. and {Elgered}, G.},
       volume = {23},
        month = nov,
        pages = {235-237},
       adsurl = {https://ui.adsabs.harvard.edu/abs/2017evga.conf..235B}
}

@article{raymond2024,
doi = {10.3847/1538-3881/ad5bdb},
url = {https://dx.doi.org/10.3847/1538-3881/ad5bdb},
year = {2024},
month = {aug},
publisher = {The American Astronomical Society},
volume = {168},
number = {3},
pages = {130},
author = {Alexander W. Raymond and Sheperd S. Doeleman and Keiichi Asada and Lindy Blackburn and Geoffrey C. Bower and Michael Bremer and Dominique Broguiere and Ming-Tang Chen and Geoffrey B. Crew and others},
title = {First Very Long Baseline Interferometry Detections at 870 um},
journal = {The Astronomical Journal}
}

@article{
venka2020,
author = {V. Venkatraman Krishnan  and M. Bailes  and W. van Straten  and N. Wex  and P. C. C. Freire  and E. F. Keane  and T. M. Tauris  and P. A. Rosado  and N. D. R. Bhat  and C. Flynn  and A. Jameson  and S. Os?owski },
title = {Lense–Thirring frame dragging induced by a fast-rotating white dwarf in a binary pulsar system},
journal = {Science},
volume = {367},
number = {6477},
pages = {577-580},
year = {2020},
doi = {10.1126/science.aax7007},
URL = {https://www.science.org/doi/abs/10.1126/science.aax7007}}

@ARTICLE{Rioja2020,
  title         = "Precise radio astrometry and new developments for the next
                   generation of instruments",
  journal       = "The Astronomy and Astrophysics Review", 
  author        = "Rioja, Mar{\'\i}a and Dodson, Richard",
  volume = {28},
  number = {6},
  month         =  oct,
  year          =  2020,
  doi = {https://doi.org/10.1007/s00159-020-00126-z}
}

@article{xie2017,
author={X. Xie and R. Bouchand and D. Nicolodi and et al.},
title={Photonic microwave signals with zeptosecond-level absolute timing noise},
journal={Nature Photon.},
volume=11,
pages=44,
year=2017
}

@ARTICLE{Boehm2018,
       author = {{B{\"o}hm}, Johannes and {B{\"o}hm}, Sigrid and {Boisits}, Janina and {Girdiuk}, Anastasiia and {Gruber}, Jakob and {Hellerschmied}, Andreas and {Kr{\'a}sn{\'a}}, Hana and {Landskron}, Daniel and {Madzak}, Matthias and {Mayer}, David and {McCallum}, Jamie and {McCallum}, Lucia and {Schartner}, Matthias and {Teke}, Kamil},
        title = "{Vienna VLBI and Satellite Software (VieVS) for Geodesy and Astrometry}",
      journal = {pasp},
         year = 2018,
        month = apr,
       volume = {130},
       number = {986},
        pages = {044503},
          doi = {10.1088/1538-3873/aaa22b},
       adsurl = {https://ui.adsabs.harvard.edu/abs/2018PASP..130d4503B}
}

@ARTICLE{Deller2007,
       author = {{Deller}, A.~T. and {Tingay}, S.~J. and {Bailes}, M. and {West}, C.},
        title = "{DiFX: A Software Correlator for Very Long Baseline Interferometry Using Multiprocessor Computing Environments}",
      journal = {pasp},
         year = 2007,
        month = mar,
       volume = {119},
       number = {853},
        pages = {318-336},
          doi = {10.1086/513572},
archivePrefix = {arXiv},
       eprint = {astro-ph/0702141},
 primaryClass = {astro-ph},
       adsurl = {https://ui.adsabs.harvard.edu/abs/2007PASP..119..318D}
}

@ARTICLE{Hoak2022,
       author = {{Hoak}, Daniel and {Barrett}, John and {Crew}, Geoffrey and {Pfeiffer}, Violet},
        title = "{Progress on the Haystack Observatory Postprocessing System}",
      journal = {Galaxies},
         year = 2022,
        month = dec,
       volume = {10},
       number = {6},
          eid = {119},
        pages = {119},
          doi = {10.3390/galaxies10060119},
       adsurl = {https://ui.adsabs.harvard.edu/abs/2022Galax..10..119H}
}

@ARTICLE{Boven2026,
    author = {Boven, E. Paul and Koelemeij, Jeroen C. J. and van Tour, Chantal and Smets, Rob and González Escudero, Rodrigo and van Langevelde, Huib Jan},
    title = "{White Rabbit in radio interferometry}",
    journal = {Experimental Astronomy},
    year = 2026,
    volume = {61},
    number = {25},
    doi = {10.1007/s10686-025-10038-4} 
}

@ARTICLE{Petrov2024,
       author = {{Petrov}, Leonid and {Kovalev}, Yuri},
        title = "{The Radio Fundamental Catalogue. I. Astrometry}",
      journal = {arXiv e-prints},
         year = 2024,
        month = oct,
          eid = {arXiv:2410.11794},
        pages = {arXiv:2410.11794},
          doi = {10.48550/arXiv.2410.11794},
archivePrefix = {arXiv},
       eprint = {2410.11794},
 primaryClass = {astro-ph.IM},
       adsurl = {https://ui.adsabs.harvard.edu/abs/2024arXiv241011794P}
}

@article{rioja2015,
  title={The Power of Simultaneous Multifrequency Observations for mm-VLBI: Astrometry up to 130 GHz with the KVN},
  author={Rioja, Mar{\'\i}a J and Dodson, Richard and Jung, Taehyun and Sohn, Bong Won},
  journal={The Astronomical Journal},
  volume={150},
  number={6},
  pages={202},
  year={2015},
  publisher={IOP Publishing}
}

@article{charlot2010,
  title={The Celestial Reference Frame at 24 and 43 GHz. II. Imaging},
  author={Charlot, P and Boboltz, DA and Fey, AL and Fomalont, EB and Geldzahler, BJ and Gordon, D and Jacobs, CS and Lanyi, GE and Ma, C and Naudet, CJ and others},
  journal={The Astronomical Journal},
  volume={139},
  number={5},
  pages={1713},
  year={2010},
  publisher={IOP Publishing}
}

@article{de2023,
  title={The Celestial Reference Frame at K Band: Imaging. I. The First 28 Epochs},
  author={de Witt, Aletha and Jacobs, Christopher S and Gordon, David and Bietenholz, Michael and Nickola, Marisa and Bertarini, Alessandra and K-band Celestial Reference Frame Collaboration and others},
  journal={The Astronomical Journal},
  volume={165},
  number={4},
  pages={139},
  year={2023},
  publisher={IOP Publishing}
}

@article{hyun2025,
  title={Optical frequency comb integration in radio telescopes: advancing signal generation and phase calibration},
  author={Hyun, Minji and Ahn, Changmin and Choi, Junyong and Baek, Jihoon and Jeong, Woosong and Je, Do-Heung and Byun, Do-Young and Wagner, Jan and Heo, Myoung-Sun and Jung, Taehyun and others},
  journal={arXiv preprint arXiv:2501.05691},
  year={2025}
}

@Article{Goti2023,
  author    = {Goti, Irene and Condio, Stefano and Clivati, Cecilia and Risaro, Matias and Gozzelino, Michele and Costanzo, Giovanni A. and Levi, Filippo and Calonico, Davide and Pizzocaro, Marco},
  title     = {Absolute frequency measurement of a {Yb} optical clock at the limit of the {Cs} fountain},
  journal   = {Metrologia},
  year      = {2023},
  volume    = {60},
  number    = {3},
  month     = may,
  pages     = {035002},
  issn      = {0026-1394},
  url       = {https://dx.doi.org/10.1088/1681-7575/accbc5},
  publisher = {IOP Publishing},
}
\bibliographystyle{aasjournalv7}



\end{document}